\documentclass{aa}  

\usepackage{graphicx}
\usepackage{txfonts}

\usepackage{hyperref}
\hypersetup{
	colorlinks=true,         
	breaklinks=true,         
    citecolor=blue,     
}

\usepackage{longtable}
\usepackage{chemformula}
\usepackage{booktabs}
\usepackage{dcolumn}
\newcolumntype{d}[1]{D{.}{.}{#1}}
\usepackage{mathrsfs}
\usepackage{xspace}
\usepackage{placeins}

\newcommand{\rev}[1]{#1}
\newcommand{\revac}[1]{#1}

\begin{document}

   \title{Unveiling gas phase \ch{H2NCO} radical\rev{: Laboratory rotational spectroscopy and interstellar search toward IRAS 16293-2422}}
    \titlerunning{Unveiling gas phase \ch{H2NCO} radical}

   \author{Marie-Aline Martin-Drumel
          \inst{1}
          \and
          Audrey Coutens
          \inst{2}
          \and
          Jean-Christophe Loison
          \inst{3}
          \and
          Jes K. Jørgensen
          \inst{4}
          \and
          Olivier Pirali 
          \inst{1}
          }

   \institute{Université Paris-Saclay, CNRS, Institut des Sciences Moléculaires d’Orsay, 91405 Orsay, France\\
              \email{marie-aline.martin@universite-paris-saclay.fr}
        \and
             Institut de Recherche en Astrophysique et Planétologie, Université de Toulouse, UT3-PS, CNRS, CNES, 9 av. du Colonel Roche, 31028 Toulouse Cedex 4, France
             \and
             Université Bordeaux, CNRS, Institut des Sciences Moléculaires, 33400 Talence, France
        \and 
            Niels Bohr Institute, University of Copenhagen, Øster Voldgade 5–7, DK-1350 Copenhagen K., Denmark
             }

   \date{Received ...; accepted ...}

 
  \abstract
   {The carbamoyl radical (\ch{H2NCO}) is believed to play a central role in the ice-grain chemistry of crucial interstellar complex organic molecules as formamide and acetamide. Yet, little is known about this radical that remains elusive in laboratory gas-phase experiments.}
   {In order to enable interstellar searches of \ch{H2NCO}, we have undertaken a mandatory laboratory characterisation of its pure rotational spectrum.}
   {We report the gas-phase laboratory detection of \ch{H2NCO}, produced by H-atom abstraction from formamide, using pure rotational spectroscopy at millimetre and submillimetre wavelengths. 
   \rev{Millimetre-wave data were acquired using chirped-pulse Fourier-transform spectroscopy while submillimetre-wave ones were obtained using Zeeman-modulated spectroscopy. Experimental measurements were guided by quantum-chemical calculations at the $\omega$B97X-D/cc-pVQZ level of theory.}
   Interstellar searches for the radical have been undertaken on the Protostellar Interferometric Line Survey (PILS) towards the solar-type protostar IRAS 16293–2422. }
   {From the assignment and fit of experimental transitions up to 660\,GHz, reliable spectroscopic parameters for \ch{H2NCO} in its ground vibrational state have been derived, enabling accurate spectral predictions. No transitions of the radical were detected on the PILS survey.  The inferred upper limit shows that \ch{H2NCO} abundance is at least \revac{60} times below that of formamide \revac{and 160 times below that of HNCO} in this source; a value that is in agreement with predictions from a physico-chemical model of this young protostar.
   }
   {}

    \keywords{methods: laboratory: molecular -- techniques: spectroscopic -- astrochemistry -- ISM: abundances --
submillimeter: general}
   \maketitle
%

\section{Introduction}

Alongside isocyanic acid (HNCO) and formamide (\ch{H2NCHO}), the carbamoyl radical (\ch{H2NCO}, also known such as \ch{NH2CO} and carbamyl in the literature) is one of the simplest molecules that incorporates the four most abundant atomic elements vital for biological systems, namely, C, H, O, and N. Because it possesses the peptide bond (\ch{NH–C(=O)–}), it holds significant potential as a key molecular precursor under prebiotic and abiotic conditions \citep[e.g.,][]{poskrebyshev2017:Theoretical}.
\ch{H2NCO} is found, both experimentally and theoretically, to be of key importance in the ice-grain chemical networks of interstellar complex organic molecules \citep[iCOMs; e.g.,][]{haupa2019:Hydrogen}.
Its formation under interstellar conditions is best explained by reactions occurring on grains, namely H-atom abstraction from formamide \citep[e.g.,][]{forstel2016:Synthesis, haupa2019:Hydrogen}, H-addition to HNCO \citep[e.g.,][]{ miller1992:Theoretical, song2016:Formation}, or to a lower extent by the reaction between \ch{NH2} and CO \citep[e.g.,][]{agarwal1985:Photochemical, bredehoft2017:ElectronInduced, ligterink2018:Formation}. These sequences are incorporated into current chemical models of gas-grain chemistry \citep{garrod2022:Formation, belloche2019:Reexploring}. 
Once formed,  the \ch{H2NCO} radical is believed to be a central intermediate in the formation of many interstellar molecules, from formamide \citep{raunier2004:Tentative, haupa2019:Hydrogen} and HNCO  \citep{hubbard1975:Ultravioletgas} to larger amides and many iCOMs such as urea \citep[\ch{NH2C(O)NH2},][]{agarwal1985:Photochemical, ligterink2018:Formation, raunier2004:Tentative, slate2020:Computational}, acetamide \citep[\ch{CH3C(O)NH2}][]{ligterink2018:Formation, belloche2019:Reexploring, garrod2022:Formation, zeng2023:Amides}, propiolamide \citep[\ch{NH2C(O)C2H},][]{alonso2021:Rotational}, and acrylamide \citep[\ch{NH2C(O)C2H3},][]{kolesnikova2022:Laboratory}. 
This common chemistry also potentially explains the observed correlations of the abundances of numerous iCOMs \citep[see, e.g.,][]{bisschop2007:Hatom, lopez-sepulcre2015:Shedding,  gorai2020:Identification, ligterink2022:Prebiotic}. \rev{Additionally, besides the species aforementioned, other molecules containing the NCO bond---and that could thus be related to H$_2$NCO---have already been detected in the interstellar medium; namely the NCO radical \citep{marcelino2018:Discovery}, N-protonated isocyanic acid \ch{H2NCO+} \citep{gupta2013:Laboratory}, methyl isocyanate \ch{CH3NCO} \citep{halfen2015:Interstellar,cernicharo2016:Rigorous}, and ethyl isocyanate \ch{C2H5NCO} \citep{rodriguez-almeida2021:First}.}

In contrast to the vast literature investigating \ch{H2NCO} formation and reactivity, only a few studies have reported a direct detection of the species, all of which were condensed phase investigations. In the late 1960s, the radical was observed using electron paramagnetic resonance spectroscopy. However, some ambiguity persisted regarding the observed isomer, either a $\sigma$-electron radical \citep[hence \ch{H2NCO},][]{livingston1967:Paramagnetic, yonezawa1968:Electron, yonezawa1969:Electron} or $\pi$-electron one \citep[\ch{HNCHO},][]{smith1966:Study, fox1968:Structure}. The latter possibility was later deemed unlikely \citep{andrews1970:Structure}. 
More recently, matrix isolation experiments have enabled the observation of the infrared spectrum of the radical, produced by H-addition to HNCO \citep{pettersson1999:Photochemistry} or H-abstraction from formamide \citep{haupa2019:Hydrogen}. The vibrational band assignments were supported by quantum chemical calculations \citep{pettersson1999:Photochemistry}. Other theoretical studies have also dived into the energetics of the [\ch{C,H2,N,O}] isomeric family \citep{shapley1999:Initio} and the thermochemistry of \ch{H2NCO} \citep{shapley1999:Initio, nagy2010:HighAccuracy, poskrebyshev2015:Calculating, poskrebyshev2017:Theoretical, nguyen1996:Reaction}.

We  here report the first detection of \ch{H2NCO} in gas-phase laboratory experiments. The radical was produced by H-atom abstraction from formamide and its pure rotational spectrum was recorded at millimetre and submillimetre wavelengths, as described in Sect. \ref{sec:lab}. Spectroscopic analysis of the experimental data has enabled the construction of a robust line frequency catalogue that was used to search for the radical towards the solar-type protostar IRAS 16293–2422 (Sect. \ref{sec:results}).

\section{Laboratory methods}\label{sec:lab}

\subsection{Quantum chemical calculations}
To initiate laboratory searches for transitions of \ch{H2NCO}, quantum chemical calculations have been performed on the species.
Geometry optimisation followed by harmonic and anharmonic frequency analysis have been carried out at the $\omega$B97X-D/cc-pVQZ level of theory \citep{chai2008:Longrange,dunning1989:Gaussian, woon1993:Gaussian} providing a reliable set of spectroscopic parameters (dipole moments projections; rotational, quartic and sextic centrifugal distortion, spin-rotation, and hyperfine constants). Special care was taken to ensure that the calculations of vibration-rotation quantities were performed in the principal axis system, as recommended in \citet{mckean2008:Scaled}.
Using Bayesian corrected rotational constants---following \citet{lee2020:Bayesian}---the spectral predictions are expected to be very close to the experimental features, as previously seen in \citet{buchanan2021:PhC3N} and \citet{Martin-Drumel2023:NBD}. 
All calculations were performed using the \textsc{Gaussian 16} suite of electronic structure programs \citep{Gaussian2016}.

\subsection{Synthesis of \ch{H2NCO}}

The \ch{H2NCO} radical was produced by H-abstraction from formamide ($\geq 99.5$\,\%, Merck), a reaction initiated by F atoms themselves produced by a 50\,W microwave discharge in \ch{F2} (5\,\% in He, Air Liquide). 
The process is similar to the one used in our previous works on \ch{CH2OH}, \ch{CH2CN}, and \ch{CH2CHO} radicals \citep{Chitarra2020, Coudert2022:CH2OH, chitarra2022:CH2CN, chahbazian2024:ch2cho}. 
The two reaction cells used to synthesise and probe the radical have been presented in detail in \citet{chitarra2022:CH2CN}: both are 50\,mm diameter cells mainly made of Pyrex, each adapted to one of our two spectrometers. The cell used for millimetre-wave chirped-pulse measurements is of about 70\,cm total length and equipped with one inlet for F-injection while the one used for submillimetre-wave measurements is about 2\,m-long and equipped with three F-injection inlets.
Using both cells, measurements were performed in a continuous flow ensured by a roots blower (EH250, Edwards) backed-up by a chemically graded pump (PFPE-E2M28, Edwards).
Since a second potential primary product of the reaction of H-abstraction from formamide is the HNCHO radical, if abstraction occurs on the nitrogen atom instead of the carbon one, initial searches for the signature of both radicals were performed using various pressure ratio of formamide and \ch{F2:He} mixture, the best signal of the \ch{H2NCO} radical was obtained using 5\,\textmu bar of formamide for a total pressure of 11\,\textmu bar. It is worth noting that the quoted pressure are only indicative as the pressure gauge was placed above the roots blower using a DN16 inlet, far downstream the probed reaction cell in order to preserve it from reactive products. Under these conditions, the pressure was similar with and without discharge. 

\subsection{Chirped-pulse millimetre-wave spectroscopy \label{sec:cp}}

A chirped-pulse Fourier-transform millimetre-wave (CP-FTMMW, Brightspec) instrument was used to perform measurements in the 75--110 GHz spectral region (W-band). Spectra were acquired using a segmented approach implemented in the high dynamic range mode of the Brightspec Edgar acquisition software, as previously detailed in \citet{chitarra2022:CH2CN}.
Initial searches were performed for both \ch{H2NCO} and HNCHO radicals using 500,000 averaged FIDs (about 1\,hour of acquisition) obtained using a pulse length of 0.5\,\textmu s. 
All measurements were repeated under the same conditions after placing a permanent magnet below the volume of the cell where the precursors interact, allowing us to identify transitions affected by the magnetic field.
Once signal of \ch{H2NCO} was found, multiple spectra were acquired under conditions found to maximise the production of this radical (5\,\textmu bar of formamide and 6\,\textmu bar of F$_2$/He, 0.25\,\textmu s excitation pulse). Free induction decay (FID) signals were acquired over 4\,\textmu s, starting 0.1\,\textmu s after the end of the excitation pulse. 
To retrieve frequency domain spectra, the FIDs were Fourier transformed using a Kaiser-Bessel apodization function over the first 2\,\textmu s since Fourier-transform over shorter FIDs yields a significant gain in signal-to-noise ratio---at the expense of the spectral resolution---which, under our experimental conditions, has proven useful to detect weak features in the spectra.
The final spectrum was averaged in the frequency domain and corresponds to 2.9 million FIDs (again, a similar spectrum was recorded with a permanent magnet placed below the reaction cell).
The resulting line frequency accuracies are estimated to be \rev{100\,kHz for strong, isolated lines and 200\,kHz for weaker and/or broader ones.}

\subsection{Zeeman-modulated submillimetre-wave spectroscopy}

The Zeeman-modulated submillimetre-wave spectrometer used in this study, which is an evolution from the double-modulation spectrometer we used previously \citep{chitarra2022:CH2CN}, has been presented in detail in \citet{chahbazian2024:ch2cho}. 
A radio-frequency synthesizer (Rhode \& Schwarz), whose 10 MHz reference signal is provided by a Rubidium clock (Stanford Research), feeds a frequency multiplication chain (Virginia Diode Inc.) operating nominally over the 75--900\,GHz range.
In this study, measurements were performed from 140\,GHz to 660\,GHz using a single path configuration in which the radiation is collimated by a teflon lens (100\,mm focal length), transmitted through the reaction cell, and focused onto the detector using either an identical teflon lens or, above 330\,GHz, an off-axis parabolic mirror (300\,mm focal length).
Schottky diode detectors (VDI) were used in the 140--220\,GHz and 220--330\,GHz spectral ranges and a mechanically-cooled Indium antimonide (InSb) bolometer (QMC) was employed above 330\,GHz. Additionally, in the 140--220\,GHz, the output radiation from the VDI chain was attenuated by about 30\,\% (Elmika, WR5 VA-015E) which was found to minimise the strong Fabry-Perot effect on the baseline without affecting the signal-to-noise ratio of the transitions.
To operate under Zeeman modulation conditions, a 16.9\,kHz sinusoidal signal---generated by a waveform generator followed by an audio amplifier---is circulating in a coil surrounding the reaction cell, resulting in an alternating magnetic field of about 3.5\,Gauss.
The signal received by the detector is then demodulated by a lock-in amplifier at the frequency of this sinusoidal signal.

The pressure conditions found to maximise the signal of the \ch{H2NCO} radical using chirped-pulse spectroscopy were found to also yield to the maximum of signal in  Zeeman-modulated spectroscopy.
Surveys around the regions where clusters of lines of \ch{H2NCO} were expected have been recorded. Thanks to the Zeeman-modulation settings, broad spectral windows, from 100\,MHz to 10\,GHz, were scanned in a straightforward fashion, without being affected by the signal of the precursor and the saturation effects it would have induced onto the lock-in amplifier in frequency-modulated or double-modulation spectroscopy, as shown already in \citet{chahbazian2024:ch2cho}. The spectra were acquired using frequency steps of 100\,kHz below 440\,GHz, and 200\,kHz above that value.
Below 330\,GHz and above 440\,GHz, each frequency point was averaged between 200\,ms and 400\,ms. In the 330--440\,GHz where the signal-to-noise ratio on \ch{H2NCO} was the strongest (a combination of Boltzmann population, power of the source, sensitivity of the bolometer, and amplitude of the Zeeman splitting with respect to the width of the line), 50\,ms averages were enough to reach an excellent signal-to-noise ratio for most lines (20 to 50). 
Finally, dedicated searches for weak $b$-type transitions (see section \ref{sec:spectro}) were performed in the 330--660\,GHz region using \rev{5\,s} averaging time.

It is worth mentioning that the profile of many \ch{H2NCO} lines (in particular $K_a=2$ ones) appears severely impacted by the effect of the magnetic field on the spin-rotation structure which results in some very asymmetrical profiles. This became particularly noticeable above 330\,GHz were the signal-to-noise ratio is the strongest. We found that we could minimise this effect by positioning a polarising grid (Pure Wave Polarizers) right after the radiation source in order to filter out the small portion of non-vertically polarised radiation emitted. A second polarising grid, also selecting vertical polarisation, was placed in front of the bolometer but its effect was not as significant, presumably because the bolometer sensor is not very sensitive to the polarisation of the radiation. A conservative frequency uncertainty has been used for transitions exhibiting asymmetric profiles. Overall, these uncertainties range from 50\,kHz (from intense, symmetric lines) to 300\,kHz (for weak, broad, or asymmetric lines) over the entire covered spectral range.

\section{Results and discussion}\label{sec:results}

\subsection{Spectroscopic considerations \label{sec:spectro}}

\begin{figure}
    \centering
    \includegraphics[width=0.8\columnwidth]{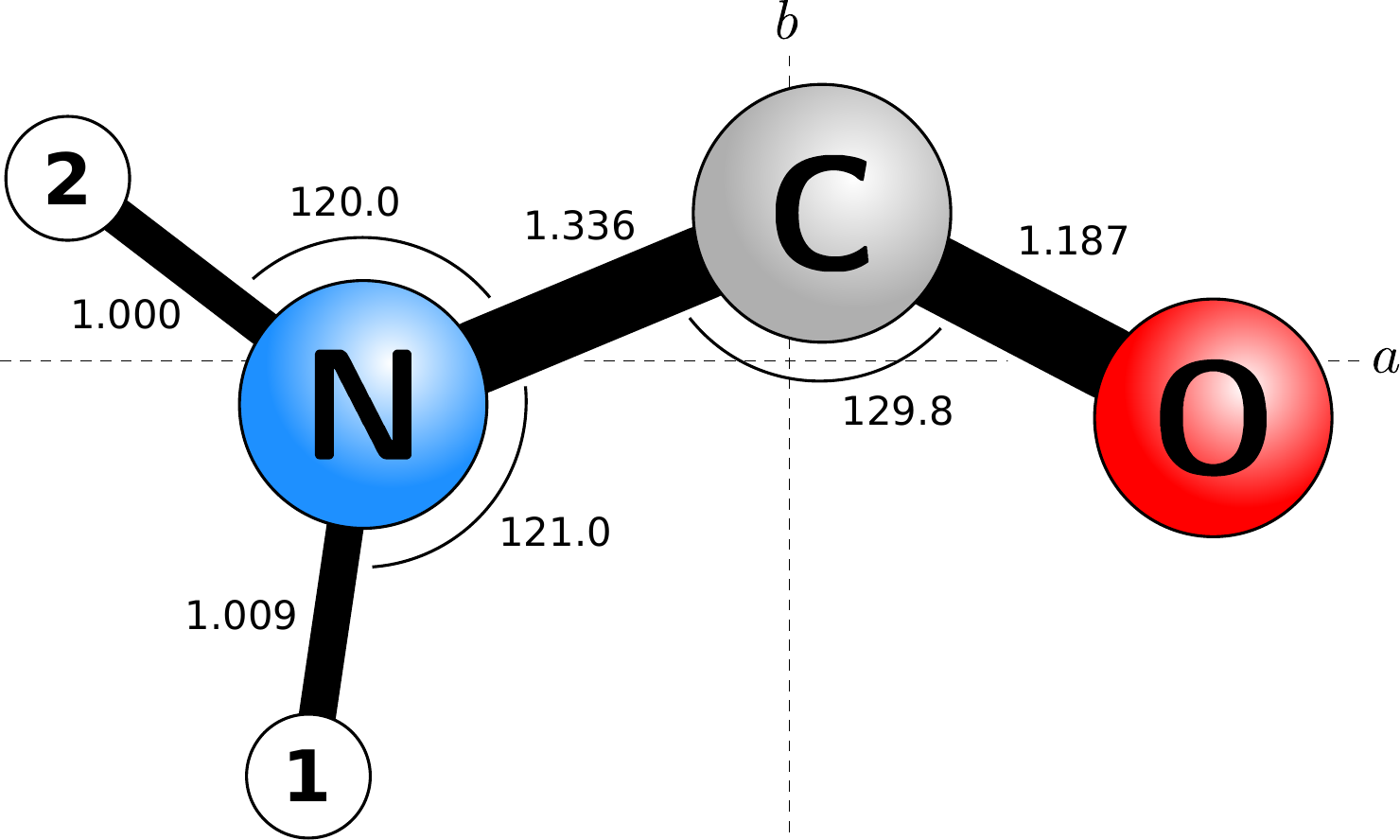}
    \caption{Equilibrium structure of \ch{H2NCO} calculated at the $\omega$B97X-D/cc-pVQZ level of theory. Geometrical parameters (bond lengths in angstroms and angles in degrees) and principal axes of inertia are reported. The figure was generated using the PMIFST software from the PROSPE collection of programs developed by \citet{kisiel2001:Assignment}. \textbf{1} and \textbf{2} refer to the H$_1$ and H$_2$ labelling used in the text.}
    \label{fig:struct}
\end{figure}

The \ch{H2NCO} radical is planar and belongs to the $C_s$ point group of symmetry. 
The molecular structure and geometrical parameters calculated in this study are displayed in Fig.~\ref{fig:struct}. 
The calculated bond lengths and angles are in good agreement with those reported in the literature \citep{pettersson1999:Photochemistry, shapley1999:Initio}, although we suspect that there was some exchange in the values pertaining to each H atom in Fig. 8 of \citet{pettersson1999:Photochemistry}.
The radical possesses dipole moment projections of 3.7\,Debye along the $a$-axis of symmetry and 0.5\,Debye along the $b$-one.

\ch{H2NCO} possesses a $^2A'$ electronic ground state and nine normal modes of vibration following the irreducible representation $\Gamma = 7 A' + 2 A''$ (see Appendix~\ref{app:calc} for vibrational frequencies and $\alpha$ values in the vibrationally excited states). 
It is a near-prolate asymmetric-top rotor with a Ray's asymmetry parameter $\kappa=-0.98$. 
Due to the presence of an unpaired electron ($S=1/2$) and three atoms with non-zero nuclear spin (N, H$_1$ and H$_2$; with $I_\mathrm{N}=1$ and $I_\mathrm{H_1}=I_\mathrm{H_2}=1/2$), each rotational energy level of $N_{K_a,K_c}$ quantum numbers is split by electron spin-rotation coupling (into a fine structure of $J$ sub-levels) and hyperfine couplings (with hyperfine levels labelled using $F_1, F_2, F$ quantum numbers). For all non-zero nuclear spin atoms, the hyperfine structure results from coupling between the electron spin and nuclear spin. Additionally, for the $^{14}$N nucleus, there is also nuclear quadrupole coupling. Because the two protons are non-equivalent, this leads to a maximum of 24 sub-levels following the coupling scheme $\mathbf{J}=\mathbf{N}+\mathbf{S}$, $ \mathbf{F_1} = \mathbf{J} + \mathbf{I}_{\ch{N}}$, $\mathbf{F_2} = \mathbf{F_1} + \mathbf{I}_{\ch{H1}}$ and $\mathbf{F} = \mathbf{F_2} + \mathbf{I}_{\ch{H2}}$. 

In the present work, we observed some transitions presenting a fully resolved hyperfine structure, the effective Hamiltonian used to model the energy levels of the vibronic ground state is thus:
$\mathscr{H} = \mathscr{H}_\mathrm{rot} + \mathscr{H}_\mathrm{sr} + \mathscr{H}_\mathrm{hfs,N}  + \mathscr{H}_\mathrm{hfs,H_1} + \mathscr{H}_\mathrm{hfs,H_2}$,
where $\mathscr{H}_\mathrm{rot}$ is the Hamiltonian operator for the pure rotational energy, $\mathscr{H}_\mathrm{sr}$ that for the electron spin-rotation, $\mathscr{H}_\mathrm{hfs,N}$ is the hyperfine Hamiltonian for the $^{14}$N nucleus (and accounts for nuclear magnetic and nuclear quadrupole interactions), and $\mathscr{H}_\mathrm{hfs,H_1}$ and $\mathscr{H}_\mathrm{hfs,H_2}$ are the hyperfine structure Hamiltonians accounting for the coupling of the spins of the two protons \ch{H1} and \ch{H2} with the spin of the unpaired electron. 
The calculated spectroscopic parameters of this effective Hamiltonian are reported in Table~\ref{tab:param}.

\begin{figure}[ht!]
    \centering
    \includegraphics[width=\columnwidth]{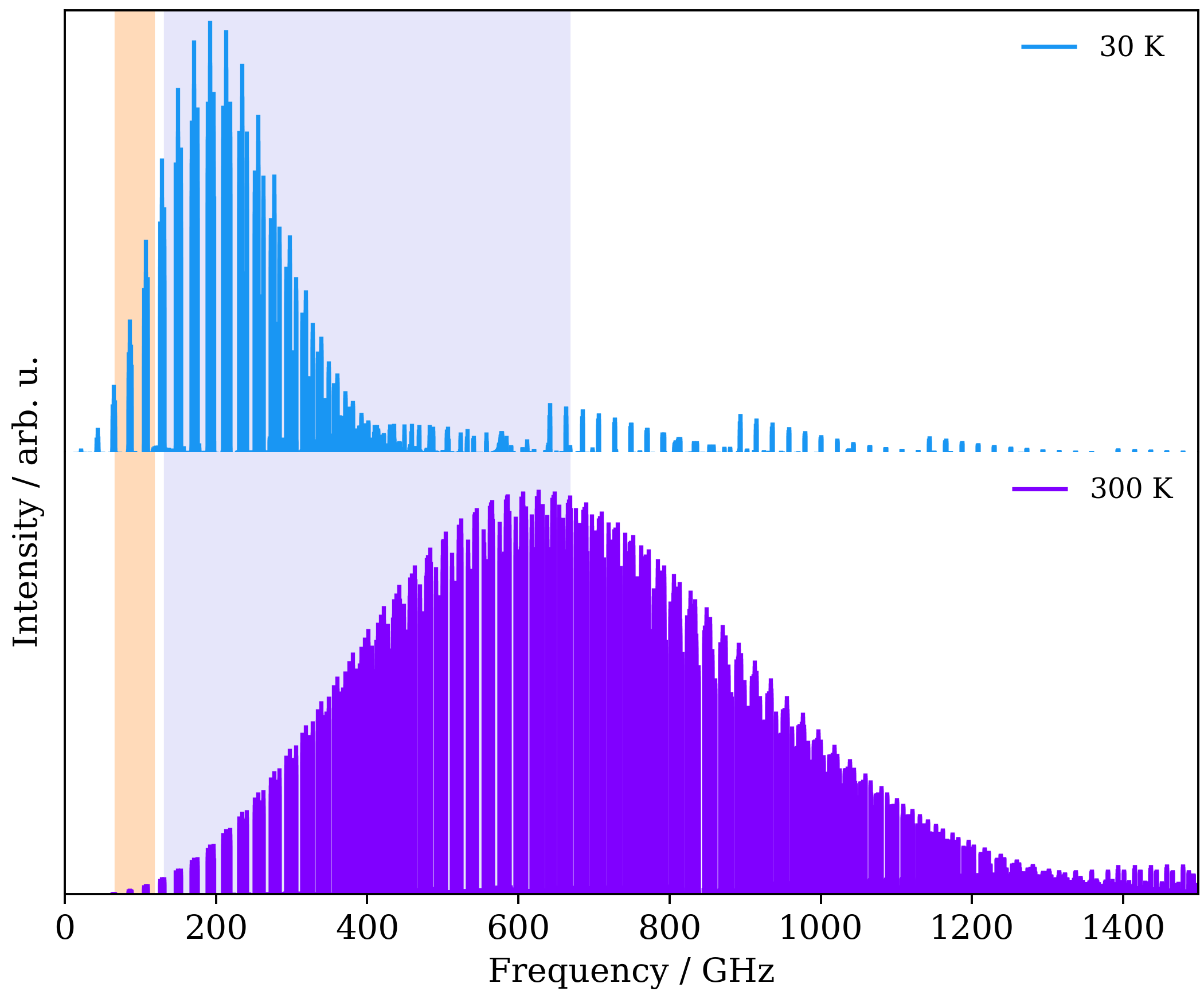}
    \caption{Predicted pure rotational spectrum of \ch{H2NCO} (in the ground vibrational state, $v=0$) up to 1.5\,THz at 30\,K (top) and 300\,K (bottom). The spectral regions investigated in this study using the CP and Zeeman-modulated spectrometers are highlighted in orange and purple, respectively.}
    \label{fig:cat}
\end{figure}

Because \ch{H2NCO} is a light species, its pure rotational spectrum extends far into the submillimetre-wave domain, even at low temperature, as illustrated on Fig.~\ref{fig:cat}. The spectral regions investigated in this study are highlighted on the figure. Due to the dipole moment projection values, $a$-type transitions dominate the spectrum; $b$-type transitions are visible in the high-frequency tail of the spectra.

\subsection{Assignments and fit}

We used the PGOPHER program \citep{western2017:PGOPHER} to search for, and assign, transitions of the \ch{H2NCO} radical. The final fit and spectral predictions were performed with the Pickett CALPGM suite of programs \citep{pickett1991:Fitting} using a Watson $S$-reduced Hamiltonian.

\begin{figure}[ht!]
    \centering
    \includegraphics[width=\columnwidth]{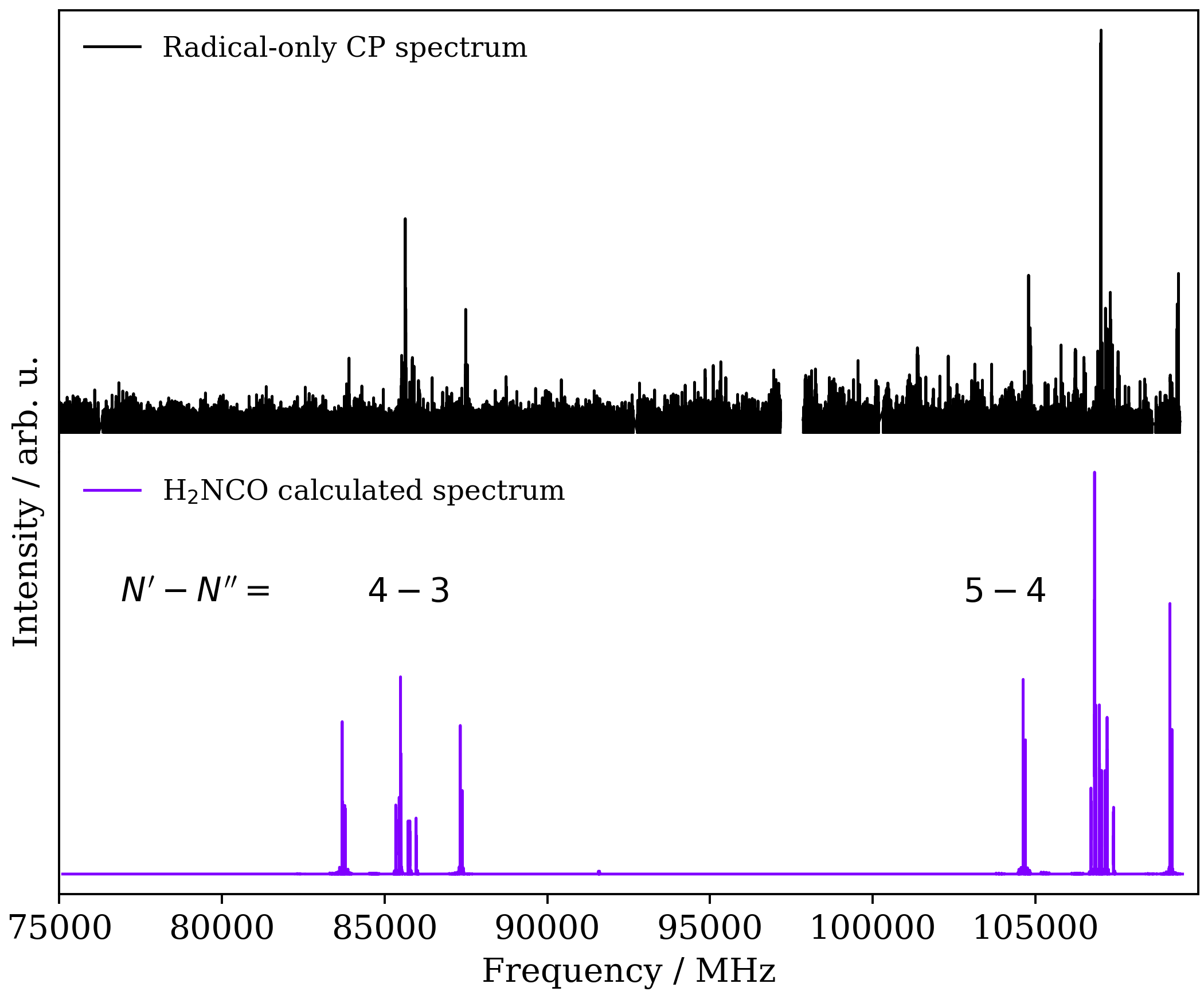}
    \caption{CP-FTMMW spectrum  after data treatment allowing to identify transitions from radical species (top trace) and comparison with predictions for \ch{H2NCO} ($v=0$, 300\,K) from the quantum chemical calculations performed in this study. The discontinuity on the experimental trace corresponds to a very noisy region that has not been plotted for the sake of clarity.}
    \label{fig:CPini}
\end{figure}

On the CP-FTMMW spectrum, two clusters of transitions closely matching what expected for the $N'-N''=4-3$ and $5-4$ transitions of \ch{H2NCO} are readily observed (Fig.~\ref{fig:CPini}). The detailed procedure used to retrieve a spectrum displaying only transitions from open-shell species and to identify these transitions is reported in Appendix~\ref{app:CPsearch}. Spectroscopic assignments of \ch{H2NCO} transitions were straightforward\rev{: the two clusters of lines are separated by about 21370\,MHz when the calculated $B+C$ value for \ch{H2NCO} is 21425\,MHz (0.3\,\% error) and the experimental transitions display the $K_a$, fine structure, and hyperfine structure patterns expected from the prediction. A closer look at the spectrum reveals that the $N'-N''=4-3$ lines and $5-4$, $K_a=0$ lines lie only about 140\,MHz and 180\,MHz higher than the prediction, respectively (about 0.2\,\% error).
}
In total 85 transitions (36 different frequencies) were assigned for the $N'-N''=4-3\ (K_a=0,1)$ and $5-4\ (K_a=0-3)$ clusters of transitions.
These initial assignments enabled preliminary adjustments of the spectroscopic constants, subsequently facilitating measurements at submillimetre wavelengths. 
A comparison of a portion of the CP spectrum with a simulation obtained using the final set of spectroscopic constants is shown in Appendix~\ref{app:CPsearch}.

\begin{figure*}[ht!]
    \centering
    \includegraphics[width=\linewidth]{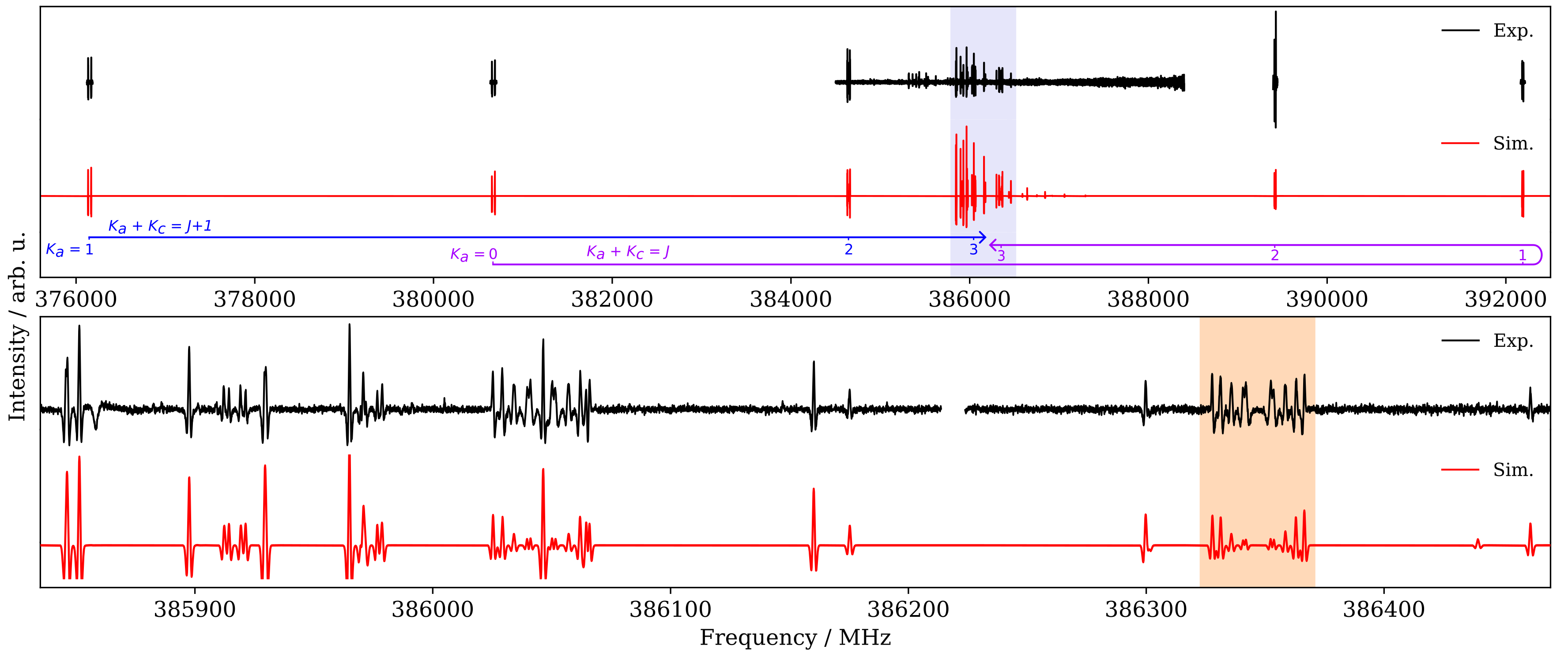}
    \caption{Overview of the $N'-N'' = 18 - 17$ $a$-type transition of \ch{H2NCO} observed experimentally and comparison with a 300\,K simulation obtained using the final spectroscopic constants (Table~\ref{tab:param}). The lowest $K_a$ values are reported for both asymmetric components on the upper panel (in blue and purple). The lower panel correspond to a zoom into the $K_a=3-12$ region highlighted in light purple in the upper panel. The area highlighted in orange displays partially resolved hyperfine structure ($K_a=3$). The simulation has been obtained using a Gaussian profile with a full-width-at-half-maximum of 1.3\,MHz; the second derivative of the PGOPHER trace is plotted for visual comparison with the experimental spectrum.}
    \label{fig:broad385}
\end{figure*}

From 146\,GHz to 660\,GHz, 3469 $a$-type transitions (1395 different frequencies) of \ch{H2NCO} have been assigned. 
An example of the $N'-N'' = 18 - 17$ $a$-type transitions, that spread over about 16\,GHz, is shown in Fig.~\ref{fig:broad385}. Spectral acquisitions were performed mostly over small regions around the predictions.
Under our experimental conditions, transitions measured using Zeeman-modulation present mostly a profile close to a second-derivative of a Gaussian line.
Many of the observed transitions display a partially resolved hyperfine structure, as for example the $K_a=3$ cluster highlighted in orange on Fig.~\ref{fig:broad385}. Further information about the hyperfine structure and how it was included in the fit is reported in Appendix~\ref{app:details}.
Fig.~\ref{fig:broad385} also illustrates the anomalously strong intensity of some $K_a$ components compared to what expected from the simulation (for instance the $K_a=2$, $K_a+K_c=J$ cluster around 389400 MHz) and asymmetry of some components  (for instance the lowest and highest frequency transitions of the $K_a=3$ cluster highlighted in orange). These effects, that may reflect high sensitivity of the spin-rotation structure to the magnetic field, are discussed in greater detail in Appendix~\ref{app:profile}. 
Finally, we observe on the experimental spectrum some transitions that likely arise from vibrational satellites of \ch{H2NCO}, as for example the small cluster of transitions around 385\,GHz not reproduced by the simulation on Fig.~\ref{fig:broad385}.
No definite quantum number assignments were performed due to the limited number of observed features (since most of these satellites lie outside the spectral windows scanned in this study). This is explained in greater details in Appendix~\ref{app:vibsat}. 

\begin{figure}[ht!]
    \centering
    \includegraphics[width=\columnwidth]{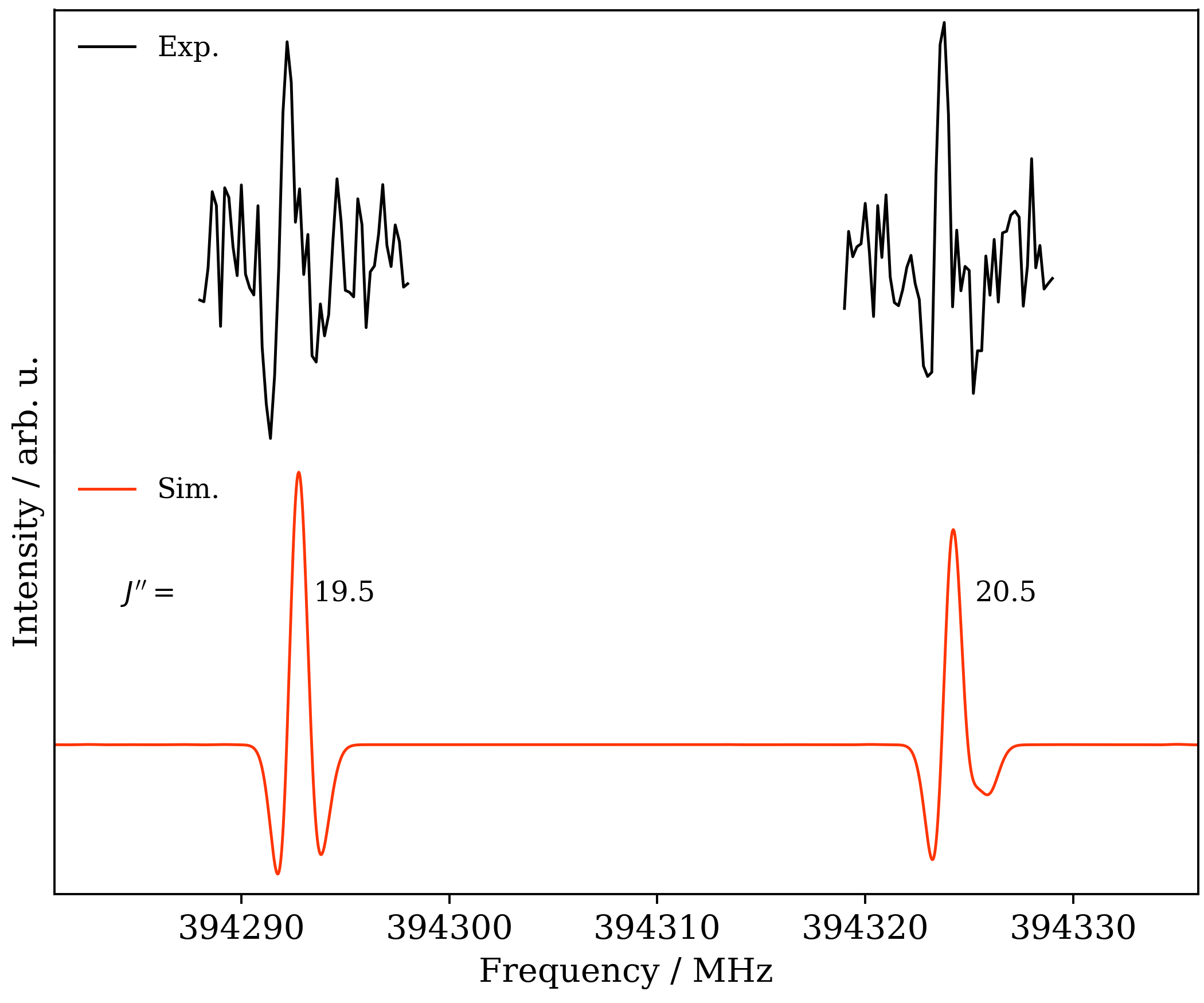}
    \caption{The $N'_{K'_a,K'_c}-N''_{K''_a,K''_c}=21_{0,21}-20_{1,20}$ $b$-type transition of \ch{H2NCO} showing two spin-rotation components. The simulation has been obtained using the final spectroscopic constants (Table~\ref{tab:param}), a 300\,K temperature, and a Gaussian profile with a full-width-at-half-maximum of 1.3\,MHz; the second derivative of the PGOPHER trace is plotted for visual comparison with the experimental spectrum. }
    \label{fig:b394}
\end{figure}

Once enough $a$-type transitions were included into the model to ensure its robustness, dedicated searches for weak $b$-type transitions were undertaken. In total, 34 lines (10 different frequencies) with $K_a'-K_a'' =0-1$ are included in the fit. An example of two such $b$-type transitions, a spin-rotation doublet, is shown in Fig.~\ref{fig:b394}. The asymmetry of the simulated transitions (in particular the higher frequency one) and to some extent of the experimental lines (despite their limited signal-to-noise ratio) is due to the presence of unresolved hyperfine components on the blue side of each transitions, hence creating a weak shoulder on the resulting lines.
The inclusion of these $b$-type transitions in the fit allows for a significant improvement in the determination of the $A$ and $\Delta_K$ constants.

Overall, 3588 lines (1441 different frequencies) with $J'' \leq 30$ and $K_a'' \leq 12$ have been assigned and included in the fit. The model contains 15 rotational parameters (rotational and centrifugal distortion constants), 6 spin-rotation parameters (including two higher order ones) and 15 hyperfine parameters. To reproduce the observed frequencies at their experimental accuracy, 22 of these parameters have been adjusted while the others were kept fixed at their calculated values. The fit takes a root-mean-square value of 107\,kHz and a reduced standard deviation of 0.84. 
The resulting set of parameters is reported in Table~\ref{tab:param} where it is compared to the calculated values. All the adjusted parameters are in excellent agreement with the calculations, with the rotational constants showing relative difference of 0.2\,\% or less from the Bayesian-scaled predicted values, and the diagonal quartic centrifugal distortion constants within about 5\,\% of the calculated values. The experimental spin-rotation and hyperfine parameters are not as close to the calculations but close enough that spectral assignments were straightforward.

\begin{table}
    \centering \scriptsize
    \caption{Spectroscopic parameters of \ch{H2NCO} in the $S$-reduction and relevant fit information. Calculated parameters are compared with those derived from the fit of the experimental measurements. 1$\sigma$ error are reported within parentheses in the unit of the last digit; they have been formatted using the PIFORM software developed by \citet{kisiel2001:Assignment}.}
    \label{tab:param}\footnotesize
    \begin{tabular}{ll d{5.7}d{5.8}}
\toprule
Parameter         & units & \multicolumn{1}{c}{Calculated $^a$} & \multicolumn{1}{c}{Exp. $v=0$}  \\
\midrule
$A$                                 & /MHz &   126337.24      &   126264.599(88)        \\
$B$                                 & /MHz &   11145.76       &    11167.74476(52)      \\
$C$                                 & /MHz &   10242.17       &    10256.93328(49)      \\
$D_N$                               & /kHz &   8.043          &        8.21279(66)      \\
$D_{NK}$                            & /kHz &   -237.9         &     -225.0451(79)       \\
$D_K$                               & /kHz &   9604.          &    10156.(86)           \\
$d_1$                               & /kHz &   -1.190         &       -1.20138(23)      \\
$d_2$                               & /kHz &   -0.05559       &       -0.06736(29)      \\
$H_N$                               & /Hz  &   0.01449        &        0.01410(37)      \\
$H_{NK}$                            & /Hz  &   -0.7904        &       -0.8267(56)       \\
$H_{KN}$                            & /Hz  &   -54.96         &      -52.140(56)        \\
$H_K$                               & /Hz  &   2201.          &    [2201.]              \\
$h_1$                               & /Hz  &   0.005175       &      [ 0.005175]        \\
$h_2$                               & /Hz  &   0.0004652      &      [ 0.0004652]       \\
$h_3$                               & /Hz  &   0.0001187      &      [ 0.0001187]       \\
\\                                                                 
$\epsilon_{aa}$                     & /MHz &   978.7          &      871.443(70)        \\
$\epsilon_{bb}$                     & /MHz &   10.52          &        4.405(10)        \\
$\epsilon_{cc}$                     & /MHz &   -42.69         &      -40.195(10)        \\
$(\epsilon_{ab}+\epsilon_{ba})/2$   & /MHz &                  &      -16.207(51)        \\
${D_N}^s$                           & /kHz &                  &        0.0572(60)       \\
${D_K}^s$                           & /kHz &                  &       -189.3(10)       \\
   \\                                                              
$a_F($N$)$                          & /MHz &   82.37          &       82.888(55)        \\
$3/2T_{aa}($N$)$                    & /MHz &   9.089          &        8.84(22)         \\
$1/4(T_{bb}-T_{cc})($N$)$           & /MHz &   0.9333         &      [ 0.9333]          \\
$T_{ab}($N$)$                       & /MHz &   -4.288         &      [-4.288]           \\
$3/2\chi_{aa}($N$)$                 & /MHz &   3.255          &      [ 3.255]           \\
$1/4(\chi_{bb}-\chi_{cc})($N$)$     & /MHz &   1.685          &      [ 1.685]           \\
$\chi_{ab}($N$)$                    & /MHz &   -0.3841        &      [-0.3841]          \\
\\                                                                 
$a_F($H$_1)$                        & /MHz &   76.73          &       79.863(96)        \\
$3/2T_{aa}($H$_1)$                  & /MHz &   3.063          &      [ 3.063]           \\
$1/4(T_{bb}-T_{cc})($H$_1)$         & /MHz &   3.153          &      [ 3.153]           \\	
$T_{ab}($H$_1)$                     & /MHz &   -5.060         &      [-5.060]           \\
\\                                                                           
$a_F($H$_2)$                        & /MHz &   -1.067         &      -2.28(19)          \\
$3/2T_{aa}($H$_2)$                  & /MHz &   24.68          &      24.93(59)          \\
$1/4(T_{bb}-T_{cc})($H$_2)$         & /MHz &   0.7960         &     [ 0.7960]           \\	
$T_{ab}($H$_2)$                     & /MHz &   1.038          &     [ 1.038]            \\
\midrule

\multicolumn{2}{l}{$\# / n\,^{b}$}
& \multicolumn{1}{c}{}
& \multicolumn{1}{c}{$3588 / 1441$}
\\
\multicolumn{2}{l}{$N'_\mathrm{max},\,K'_{a\,\mathrm{max}}$}
& \multicolumn{1}{c}{} 
& \multicolumn{1}{c}{$31,\, 12$} 
\\
\multicolumn{1}{l}{RMS $\,^c$} & /MHz
& \multicolumn{1}{c}{}
& \multicolumn{1}{c}{0.107} 
\\
\multicolumn{2}{l}{$\sigma\,^{d}$}
& \multicolumn{1}{c}{}
& \multicolumn{1}{c}{0.84} 
\\
\bottomrule     
    \end{tabular}
    
\smallskip

\begin{minipage}{1\columnwidth}  \footnotesize
    $^a$ The calculated $A, B, C$ rotational constants have been Bayesian corrected by a 0.9866 factor \citep[see][]{lee2020:Bayesian}, all the other calculated parameters are equilibrium values.\\    
    $^b$ Number of lines ($\#$) and number of different frequencies ($n$) \\
    $^c$ Root mean square value\\
    $^d$ Weighted standard deviation
\end{minipage}
\end{table}

Using the spectroscopic parameters of \ch{H2NCO} derived from this work, reliable predictions can be made up to about 700\,GHz (extrapolation to higher frequencies than the experimental measurements can be unreliable, thus prediction above 700\,GHz using the present model should be taken with caution). 
A spectral catalogue at 300\,K \rev{(with fully resolved hyperfine structure)} is available in the electronic supplementary material together with the assignment and fit files.
\rev{
The catalogue frequencies should be accurate to 100\,kHz or better up to 650\,GHz, i.e. 1.0\,km\,s$^{-1}$ at 30\,GHz, 0.6\,km\,s$^{-1}$ at 50\,GHz, 0.3\,km\,s$^{-1}$ at 100\,GHz, and 0.1\,km\,s$^{-1}$ at 350\,GHz. It is worth noting that these values may be larger than the linewidths observed in cold interstellar sources below 50\,GHz. For instance, TMC-1 linewidths are 0.72\,km\,s$^{-1}$ around 40 GHz in the QUIJOTE survey \citep{agundezl2022:Propargyl}. Future experimental work targeting the lowest frequency $K_a=0$ transitions of \ch{H2NCO} ($N=1-0$ around 21.4\,GHz, $N=2-1$ around 42.8\,GHz, and  $N=3-1$ around 64.2\,GHz) may help reduce these uncertainties. To help recalculate the catalogue intensities at various temperatures, partition} functions have been calculated (from 300\,K to 9\,K) and are reported in Appendix  \ref{app:partFunc}.

As final notes, in the course of the present measurements, we observed transitions of carbamyl fluoride (\ch{FC(O)NH2}) in the 75--110\,GHz region (no higher frequency transitions have been measured due to the open-shell species selectivity of the Zeeman-modulation measurements). The species was previously investigated only in the 16--40\,GHz region by \citet{rigden1966:Microwave}; the present measurements  and combined fit with the literature data are reported in Appendix~\ref{app:fconh2}.
Lastly, we did not observe transitions that could belong to the HNCHO isomer of \ch{H2NCO}. \rev{This non-observation can be rationalised by several factors: 1) the lower intensity of HNCHO transitions compared to \ch{H2NCO}; 2) the fact that  HNCHO may not be synthesised in our experiment; and/or 3) the fact that the transitions of HNCHO could lie outside of the spectral ranges scanned in this study. With respect to point 1, from the spectroscopic information given by the quantum chemical calculations performed in this work (see Appendix~\ref{app:hncho}), the lowest energy \textit{cis} conformer of HNCHO possesses dipole moment projections lower than 1 D. Would \textit{cis}-HNCHO be synthesised in concentration similar to \ch{H2NCO}, its transitions may thus remain below our detection limit. The \textit{trans} conformer of HNCHO possesses more favourable dipole moment projections (over 2 Debye) but  lies 10.5 kJ.mol$^{-1}$ higher in energy than the \textit{cis}. If thermodynamic factors govern the conformer population under our experimental conditions, much lower concentration of this species may be found, yet again impacting its detectability.
Concerning point 2, previous studies have indicated that even though HNCHO should be the most stable isomer, H-abstraction would favour \ch{H2NCO} \citep{zeng2023:Amides}. Regardless, some experimental tests could be performed by using Cl atoms instead of F ones for the H-abstraction, to thus assess whether the halogen nature plays a role in the favoured site for H-abstraction on the precursor (carbon versus nitrogen atom).
Finally, concerning point 3, a previous study has reported the existence of a fast non-planar transition between the \textit{cis} and \textit{trans} conformations of the HNCHO radical that leads to an equilibrium between these configurations \citep{andrews1970:Structure}. Further high level quantum chemical calculations may be warranted to confirm that point and determine the actual equilibrium geometry of HNCHO, and thus the best predicted rotational constants to perform dedicated search for the species in specific spectral regions.
}

\subsection{Astrochemical implications}
To test the new spectroscopic predictions for \ch{H2NCO}, we carried out a search for it using the data from the Protostellar Interferometric Line Survey (PILS). PILS is a large spectral survey carried out towards the solar-type protostar IRAS~16293--2422 with the Atacama Large millimetre/submillimetre Array (ALMA) 12\,m array and the Atacama Compact Array (ACA) in band 7. It has a spatial resolution of $\sim$0.5$\arcsec$ and a sensitivity of 4--5\,mJy\,beam$^{-1}$ for a bin width of 1\,km\,s$^{-1}$ \citep{Jorgensen2016}. This survey, which covers the spectral range 329--363\,GHz, has led to several new molecular detections in low-mass protostars \citep[e.g.,][]{Lykke2017,Ligterink2017,Fayolle2017,Coutens2018,Coutens2019, Manigand2021}. Formamide and several of its isotopologues (NHDCHO, NH$_2$CDO, NH$_2$$^{13}$CHO) were detected at a position located one beam offset from the continuum peak position of the B component \citep[IRAS16293~B in the following; ][]{coutens2016:ALMAPILS}. Cyanamide (H$_2$NCN) was also detected towards the same position \citep{Coutens2018}. A search for the H$_2$NCO radical is consequently pertinent towards this source.

\begin{figure}[ht!]
    \centering
    \includegraphics[width=\columnwidth]{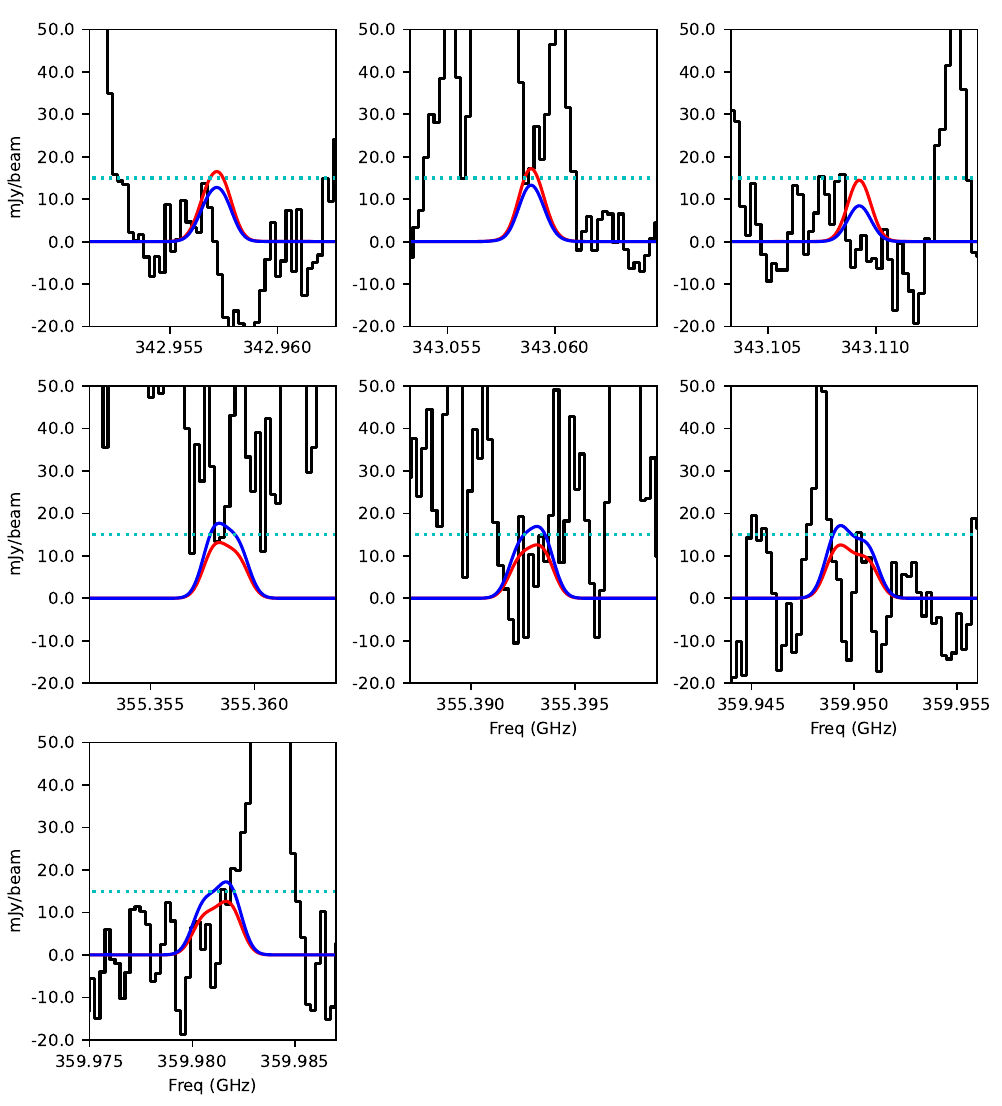}
    \caption{Undetected transitions of H$_2$NCO that were used to determine the column density upper limits. The ALMA observed spectrum of IRAS16293~B is in black. The LTE model for $T_{\rm ex}$ = 300\,K is in red, while the one for $T_{\rm ex}$ = 125\,K is in blue. \rev{The dotted cyan line corresponds to the 3 rms level per 1 km\,s$^{-1}$, 1 km\,s$^{-1}$ being the typical line width at this position.} 
    }
    \label{fig:non-detection}
\end{figure}

Local thermodynamic equilibrium (LTE) models were produced with the CASSIS\footnote{CASSIS has been developed by IRAP-UPS/CNRS \citep{Vastel2015},   \url{http://cassis.irap.omp.eu}} software and compared to the spectrum extracted at the same position as the previous molecular studies.
The models assume a linewidth of 1 km\,s$^{-1}$, a source size of 0.5$\arcsec$ and a $\varv_{\rm LSR}$ of 2.7\,km\,s$^{-1}$, \revac{and take into account the total partition function given in Appendix \ref{app:partFunc}}. The excitation temperatures of complex organic molecules vary in this source  between $\sim$125\,K and 300\,K \citep{Jorgensen2018}.  Different excitation temperatures were consequently assumed in the search, but independent on the assumed temperature no detected lines could be claimed above noise of the observed spectrum.
We used the brightest H$_2$NCO lines predicted by the models that are not contaminated by emission of other molecules to derive upper limits on the column density, \rev{namely the $a$-type, $R$-branch transitions involving $N''_{K''_a,K''_c} = 15_{5,10/11}$ (both spin-rotation components around 342.957 and 343.059\,GHz), $16_{1,16}$ (around 355.358 and 355.393\,GHz),  $16_{1,16}$ (around 359.950 and 359.981\,GHz), and $N''_{K''_a,K''_c}(J'') = 15_{6,9/10}(15.5)$ (around 343.109\,GHz). }
Based on the transitions at 342.957, 343.059 and 343.109\,GHz, the upper limit is estimated to be about \revac{1.7}\,$\times$\,10$^{14}$\,cm$^{-2}$ if the excitation temperature is similar to NH$_2$CHO ($\sim$ 300 K, \citealt{coutens2016:ALMAPILS}). For a lower excitation temperature of $\sim$ 125 K, we used the groups of transitions at 355.358, 355.393, 359.950, and 359.981\,GHz and found a very close upper limit of 8.0\,$\times$\,10$^{13}$\,cm$^{-2}$. These non-detected lines are shown in Figure \ref{fig:non-detection}.

\citet{coutens2016:ALMAPILS} found \revac{column densities of NH$_2$CHO and HNCO equal to 1.0\,$\times$\,10$^{16}$ and 2.7\,$\times$\,10$^{16}$ cm$^{-2}$, respectively, using their} $^{13}$C isotopologue lines and a $^{12}$C/$^{13}$C ratio of 68 \citep{Milam2005}.
H$_2$NCO is consequently \revac{$\gtrsim$60} times less abundant than NH$_2$CHO and \revac{$\gtrsim$160  than HNCO} towards IRAS16293~B. \revac{The radical NCO has never been searched towards IRAS~16293--2422 as far as we know. We searched for it in the PILS data but we only derived an upper limit of 7\,$\times$\,10$^{14}$ cm$^{-2}$ for $T_{\rm ex}$ = 300 K and 5\,$\times$\,10$^{14}$ cm$^{-2}$ for $T_{\rm ex}$ = 125 K}. \revac{Among the other NCO-bearing molecules detected in IRAS16293B, CH$_3$NCO has a column density of $\sim$3--4\,$\times$\,10$^{15}$ cm$^{-2}$ \citep{Ligterink2017,martin-domenech2017} and is at least 17 times more abundant than H$_2$NCO. Acetamide (CH$_3$C(O)NH$_2$) has also been tentatively detected by \citet{ligterink2018:Formation} with a column density of (9--25)\,$\times$\,10$^{14}$ cm$^{-2}$ depending on the excitation temperature. If confirmed, it would mean that this molecule is $\gtrsim$5-15 times more abundant than H$_2$NCO.}


To understand the astrophysical implications of the non-detection of \ch{H2NCO} in IRAS16293~B we compare the upper limit derived above to the calculated abundance obtained using the model used in \cite{RN11273, Manigand2021} which uses the Nautilus code \citep[][]{RN7545}---a 3-phase gas, dust grain ice surface, and dust grain ice mantle time dependent chemical model employing kida.uva.2014 \citep{RN6632} as the basic reaction network. The model was recently updated to better describe iCOMs on grains and in the gas-phase \citep[800 individual species are included in the network and are involved in 9000 separate reactions, ][]{RN11273, Manigand2021}. The physical model to describe IRAS16293~B evolution consists in the two successive evolutionary stages of a low-mass protostar: a uniform and constant stage corresponding to the pre-stellar phase, or the cold-core phase, followed (after $1\times10^6$\,years) by a collapse phase as described in \cite{Manigand2021}. The following conditions are used in the pre-stellar phase: homogeneous gas with a density equal to $2.5\times10^4$\,cm$^{-3}$; temperature of 10\,K for both gas and dust; visual extinction of 10 mag; cosmic-ray ionisation rate of $1.3\times10^{-17}$\,s$^{-1}$; standard external UV field of 1\,G$_0$; and the same initial abundances than in \cite{Manigand2021}.

\begin{figure}[ht!]
    \centering
    \includegraphics[width=\columnwidth]{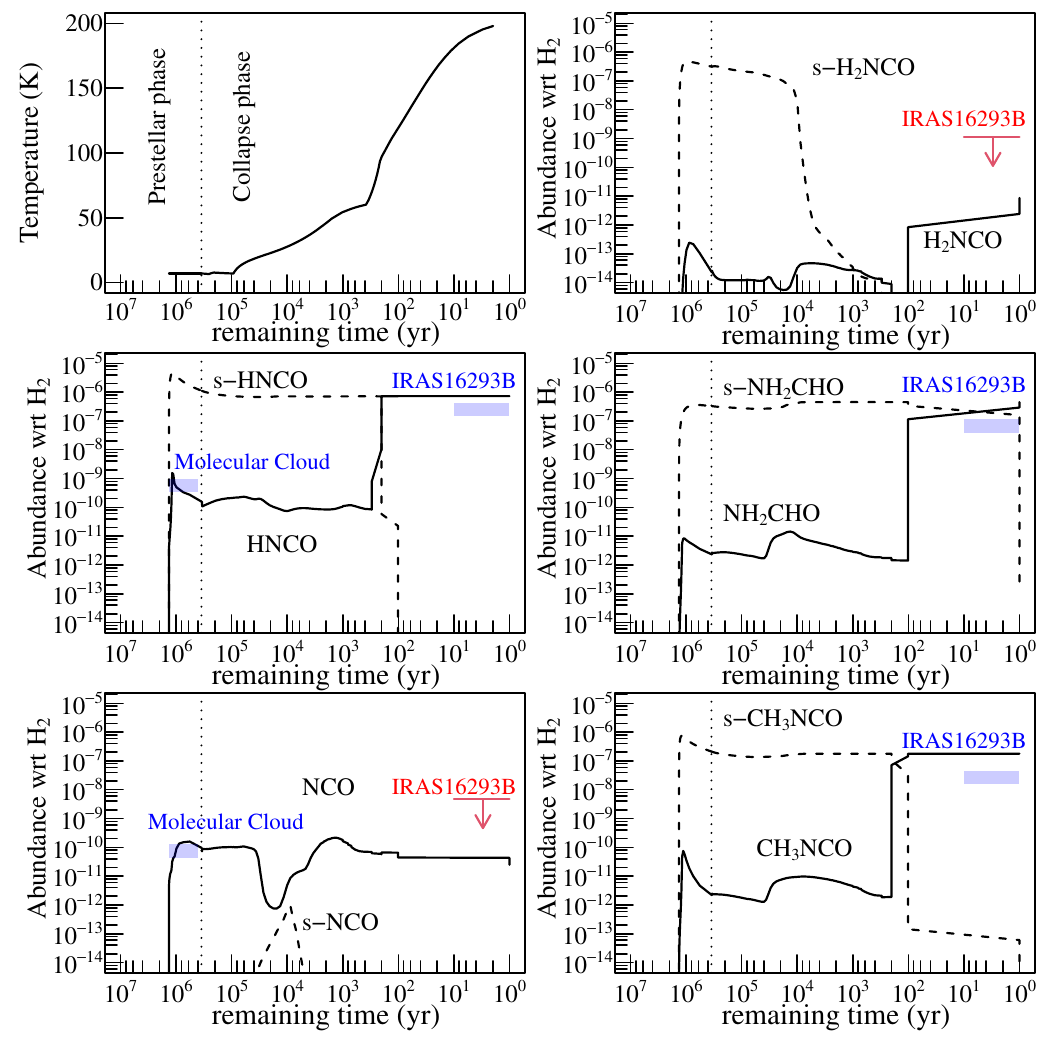}
    \caption{Time evolution of the temperature and the abundances (with respect to molecular hydrogen) in the gas phase (solid lines) and on the grain surface (dashed) of HNCO, \ch{H2NCO}, \ch{NH2CHO} \rev{NCO, and \ch{CH3NCO}} species during the cold-core phase and the collapse phase. The time axis is reversed to better visualise the abundances evolution. The observed gas-phase abundances are indicated in blue. Detection in IRAS16293~B are from \protect\cite{coutens2016:ALMAPILS, RN9736, RN11172} for HNCO and from \protect\cite{coutens2016:ALMAPILS, RN10761, RN11172} for \ch{NH2CHO}. Detection in dense molecular clouds are from \protect\cite{RN9136, RN6784} and \citet{RN6804}. The upper limit for the relative abundance of gas phase \ch{H2NCO} \rev{and NCO in IRAS16293~B} are shown in red. The observed abundances for IRAS16293~B are calculated from the observed column densities using an \ch{H2} column density equal to $1.5\times 10^{23}$\,cm$^{-2}$ so that the chemical model reproduces the observed abundance of \ch{CH3OH} (this is equivalent to using abundances relative to \ch{CH3OH}).}    
    \label{fig:JC}
\end{figure}

The evolution of the abundances during the cold-core phase and the collapse phase is shown in Figure \ref{fig:JC}. \ch{H2NCO} and the species directly related to it, HNCO and \ch{NH2CHO}, are mostly produced on grain surfaces and the closed shell species HNCO and \ch{NH2CHO} are released in the gas-phase when the temperature reaches the desorption temperature, between 120 to 130\,K---the binding energy of HNCO is measured equal to 3900\,K \citep{RN7064} and that of \ch{NH2CHO} is between 5056 and 6990\,K \citep{RN9130}. More precisely, in our network HNCO is mainly produced by the reaction s-N + s-HCO, where s- means species on grain. This reaction is effective even at 10\,K in the cold-core phase as N is, along with atomic hydrogen, relatively mobile on grains when we consider its adsorption energy equal to 700\,K \citep{RN10147, RN10829} and $E_\mathrm{diffusion} = 0.4\times E_\mathrm{binding}$ \citep{RN1878}. \ch{H2NCO} is produced by hydrogenation of HNCO (which presents a barrier but is made possible by tunnelling), H-atom abstraction from s-\ch{NH2CHO} and also by the s-CN + s-\ch{H2O} reaction, which according to a recent theoretical study is possible by a concerted effect of water molecules in the ice \citep{RN9079}. Finally, \ch{NH2CHO} is produced by the hydrogenation of \ch{H2NCO} and by the s-\ch{NH2} + s-HCO reaction. The agreement between the modelled abundances for HNCO and \ch{NH2CHO} and the observations for IRAS16293~B (and also molecular clouds for HNCO) is relatively good, given the many uncertainties and unknowns (for example the branching ratios between the s-\ch{NH2CHO} and s-\ch{NH3} + s-CO pathways for the s-H + s-\ch{H2NCO} reaction).

Figure \ref{fig:JC} also shows that the calculated abundance of s-\ch{H2NCO} is high as long as the temperature does not exceed 25\,K. Above this temperature, certain radicals become mobile on ice---s-HCO, for example, has a relatively low binding energy around 3000\,K \citep{RN7696, RN7168, RN8226, RN7750}, but not \ch{H2NCO}, whose binding energy is calculated to be around 7000\,K \citep{RN7750}. Then, as the s-radical + s-radical reaction are barrierless, the radicals will react with each other on the grains before being able to desorb. Note that as the diffusion energy of the radicals is always notably lower than the binding energy ($E_\mathrm{diffusion} = 0.4\times E_\mathrm{binding}$ in our model), the radicals will diffuse before desorbing and will therefore be rapidly destroyed on the grains before they can be injected into the gas phase. The only possibility for radicals to desorb is if they are produced at temperatures above their desorption temperature, as in the case of NO, or possibly HCO, detected in IRAS16293~B \citep{RN9862, RN9860}. The non-detection of \ch{H2NCO} suggests that this radical is not produced at high temperature on grains nor is it produced efficiently in the protostellar hot gas phase. However, its non-detection does not compromise its important role as an intermediate in \ch{NH2CHO} production.

\section{Conclusions}

We have conducted an extensive investigation into the room-temperature pure rotational spectrum of the \ch{H2NCO} radical (previously never detected in the gas phase in the laboratory) in the 75--660\,GHz region. Close to 1500 frequencies have been assigned in $v=0$, upon guidance from quantum-chemical calculations, enabling a robust set of spectroscopic parameters to be derived for the species. 
Using this model, accurate spectral predictions can be made over the entire millimetre- and submillimetre-wave spectral regions. This opens up the possibility for interstellar searches for the radical in cold to warm environments ($T\leq 300$\,K). Searches have been undertaken toward IRAS16293~B, a source where the related molecule \ch{NH2CHO} is abundant, using the PILS survey. No transition of the radical was detected and upper limits on the column densities have been derived. The absence of detection of \ch{H2NCO} in the gas phase is consistent with a physico-chemical model of the formation of complex molecules during the earliest stages of the source.

\begin{acknowledgements}
    This work has been supported by the \textit{R\'egion Ile-de-France}, through DIM-ACAV$^+$, the \textit{Agence Nationale de la Recherche} (ANR-19-CE30-0017-01), the ``\textit{Investissements d’Avenir}'' LabEx PALM (ANR-10-LABX-0039-PALM), and the \textit{Programme National} ``\textit{Physique et Chimie du Milieu Interstellaire}'' (PCMI) of CNRS/INSU with INC/INP co-funded by CEA and CNES.
    Quantum chemical calculations were performed using HPC resources from the ``\textit{M\'esocentre}'' computing centre of CentraleSupélec and \'Ecole Normale Sup\'erieure Paris-Saclay supported by CNRS and R\'egion \^Ile-de-France (\url{http://mesocentre.centralesupelec.fr/}).
    This paper makes use of the following ALMA data: ADS/JAO.ALMA\#2013.1.00278.S. ALMA is a partnership of ESO (representing its member states), NSF (USA) and NINS (Japan), together with NRC (Canada) and NSC and ASIAA (Taiwan), in cooperation with the Republic of Chile. The Joint ALMA Observatory is operated by ESO, AUI/NRAO and NAOJ. A. C. received financial support from the European Research Council (ERC) under the European Union’s Horizon 2020 research and innovation programme (ERC Starting Grant ``Chemtrip'', grant agreement No 949278). J.K.J. acknowledges support from the Independent Research Fund Denmark (grant number 0135-00123B).
    We thank T. Huet for the loan of the audio-amplifier and M. Goubet, Z. Kisiel, M. D. Marshall, and B. J. Esselman for useful discussion regarding the vibration-rotation calculations with Gaussian. 
\end{acknowledgements}

%
%
\bibliographystyle{aa} 
\bibliography{H2NCO.bib}

\begin{appendix} 

\section{Additional calculation outputs }\label{app:calc}
\FloatBarrier

\begin{table}[ht!]
    \centering
    \setlength\tabcolsep{4pt}
    \caption{Fundamental modes of \ch{H2NCO} calculated in this study ($\omega$B97X-D/cc-pVQZ). Energies are reported in cm$^{-1}$ and intensities in km\,mol$^{-1}$.}
    \begin{tabular}{crrrrc}
    \toprule
   Mode       &  $E$(harm) & $E$(anharm) & $I$(harm) & $I$(anharm) & Sym. \\ \midrule
      $\nu_1$ &   3754     & 3615        &  71.83    &  59.41      & A$'\  $\\
      $\nu_2$ &   3556     & 3401        &  11.46    &  3.85       & A$'\  $\\
      $\nu_3$ &   1895     & 1874        &  341.28   &  305.02     & A$'\  $\\
      $\nu_4$ &   1617     & 1576        &  61.12    &  39.26      & A$'\  $\\
      $\nu_5$ &   1249     & 1222        &  58.97    &  42.03      & A$'\  $\\
      $\nu_6$ &   1106     & 1099        &  0.30     &  0.05       & A$'\  $\\
      $\nu_7$ &   540      & 533         &  6.29     &  27.21      & A$'\ $\\
      $\nu_8$ &   631      & 601         &  0.47     &  0.58       & A$''$\\
      $\nu_9$ &   282      & 422         &  204.63   &  146.52     & A$''$\\ 
      \bottomrule
    \end{tabular}
    \label{tab:vib}
\end{table}

\begin{table}[ht!]
    \centering
    \caption{Calculated $\alpha$ values (in MHz) for \ch{H2NCO} in excited vibrational states ($\omega$B97X-D/cc-pVQZ). Predictions of the rotational constants values in $v_i=1$ can be retrieved using the formula $B_i = B_0 + \alpha_{B\,i}$.}
    \begin{tabular}{cd{5.1}d{5.1}d{5.1}}
    \toprule
   State      &  \multicolumn{1}{c}{$\alpha_A$} & \multicolumn{1}{c}{$\alpha_B$} & \multicolumn{1}{c}{$\alpha_C$} \\ \midrule
      $v_1=1$ &   -167.0    & -17.3       &  -14.1\\
      $v_2=1$ &   -324.0    & -12.6       &  -12.1\\
      $v_3=1$ &   -1483.3   & -33.1       &  -33.3\\
      $v_4=1$ &   816.1     & 4.7         &  -13.4\\
      $v_5=1$ &   1397.9    & -55.2       &  -51.4\\
      $v_6=1$ &   1016.2    & -5.4        &  -27.8\\
      $v_7=1$ &   -3491.4   & -14.7       &  -4.4 \\
      $v_8=1$ &   4879.1    & -1.3        &  0.1  \\
      $v_9=1$ &   -3714.2   & -23.9       &  11.6 \\ 
      \bottomrule
    \end{tabular}
    \label{tab:alphas}
\end{table}

\FloatBarrier
\section{Initial searches for radical products using the chirped-pulse spectrometer} \label{app:CPsearch}

As mentioned in section \ref{sec:cp}, we performed initial searches for H-abstraction products from formamide using the chirped-pulse spectrometer because it ideally allows to rapidly, and repeatedly, scan a broad spectral region under various experimental conditions.
The initial searches were performed using a 0.5 \textmu s excitation pulse which was found to give reasonably good signal for both strong and moderate dipole moment transitions, in order to scan without bias for potential reaction products in our experiment.
Starting with a pressure found to maximize the signal of formamide in our flow conditions (5 \textmu bar), we recorded several chirped-pulse spectra for formamide:\ch{F2}/He ratio of 1:0.5, 1:1, 1:2, and 1:3. Each of these spectra was recorded for about 1\,hour (500,000 averages) and the measurement was repeated after placing a permanent magnet below the cell. 
Before recording the spectra, we ensured that the added pressure of \ch{F2}/He did not degrade the signal of the precursor (as high pressure can cause the polarized molecules to relax faster by collision hence yielding shortened FIDs).

Because the experimental conditions were relatively stable during the scans, and because the magnet does not affect the signal of the precursor lines nor that of the close-shell products of the reaction, it was possible to essentially subtract the spectrum recorded with the magnet from the one recorded without it for each formamide:\ch{F2}/He pressure ratio. By doing so, only transitions whose intensity are affected by the magnetic field appear on the resulting spectrum. Concretely, because chirped-pulse spectra display some negative noise and spurs, and because there was some slight variations in the line intensity of the precursor, special care was taken in the subtraction procedure in order to not add positive noise to the subtracted spectrum and to filter all signal from the precursor. 
To do so, all negative intensity points on the spectrum recorded with the magnet are set to zero. Then, for each frequency point, the resulting intensity from this treated spectrum is multiplied by a constant factor (typically 1 to 1.2, to account for intensity fluctuations of the precursor signal) and subtracted from the corresponding intensity on the spectrum recorded without magnet. 
To enable a straightforward display of the subtracted spectrum, a negative intensity cut-off is applied for values below the noise level. Finally, remaining spurious signals are identified by visual inspection of the subtracted spectrum and discarded (intensity set to zero), as are some particularly noisy regions of the spectrum (typically with our instrument the region around 97.5 GHz is extremely noisy). 

\begin{figure}[ht!]
    \centering
    \includegraphics[width=\linewidth]{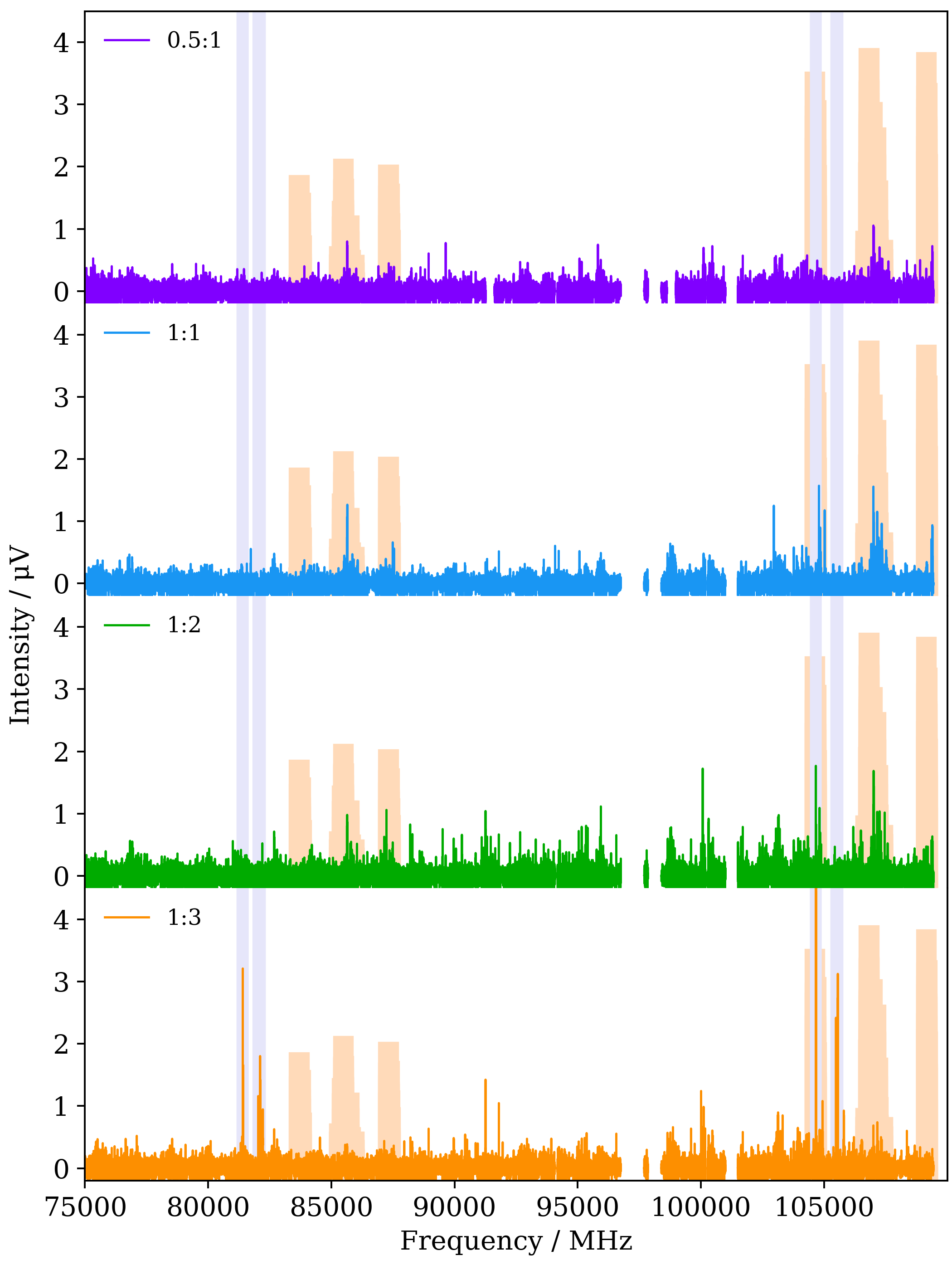}
    \caption{Comparison of the different CP spectra after data treatment allowing to identify transitions from radical species (formamide:\ch{F2}/He pressure ratio of 1:0.5, 1:1, 1:2, and 1:3, from top to bottom). The shaded areas correspond to $\pm 200$\,MHz windows around the spectral predictions from the NCO and \ch{H2NCO} radicals (in light blue and orange, respectively). For NCO, predictions are from the CDMS database \citep{endres2016:Cologne} based on the experimental work by \citet{kawaguchi1985:Microwave}. For \ch{H2NCO}, calculated values are from this work and the relative intensity is proportional to the expected line intensity at 300\,K, in arbitrary units. Discontinuities on the experimental spectra correspond to very noisy regions that have not been plotted for the sake of clarity.}
    \label{fig:CPsearches}
\end{figure}

Overall, this procedure allows us to search for transitions arising from radical species on each spectrum. Fig.~\ref{fig:CPsearches} shows the resulting spectra. Around the areas where transitions of \ch{H2NCO} are expected, features are easily distinguishable on all spectra; they are the strongest on the 1:1 and 1:2 ratio spectra. 
The only other radical clearly identifiable on the spectra is NCO, which presents strong transitions on the 1:3 ratio spectrum. \rev{The experimental data do not allow for a comparison of the production of the two radicals in terms of concentration because of both the different polarisation of the two radicals (which depends on their dipole moment projections) and the respective volume where they are present (which depends on the lifetime/reactivity of each species). We, however, qualitatively see here that a high concentration of F atoms yields multiple H-abstraction from formamide.
}
As a final note, transitions of \ch{H2NCO} are more than 1000 times weaker than transitions of formamide  on the CP spectra, thus the identification of the radical features was only possible because the precursor spectral density was not too large, allowing for transitions of \ch{H2NCO} to lie in transparency windows. The methodology used in the present study would not be applicable for a precursor presenting a very dense spectrum in the W-band region.

\begin{figure}[ht!]
    \centering
    \includegraphics[width=\linewidth]{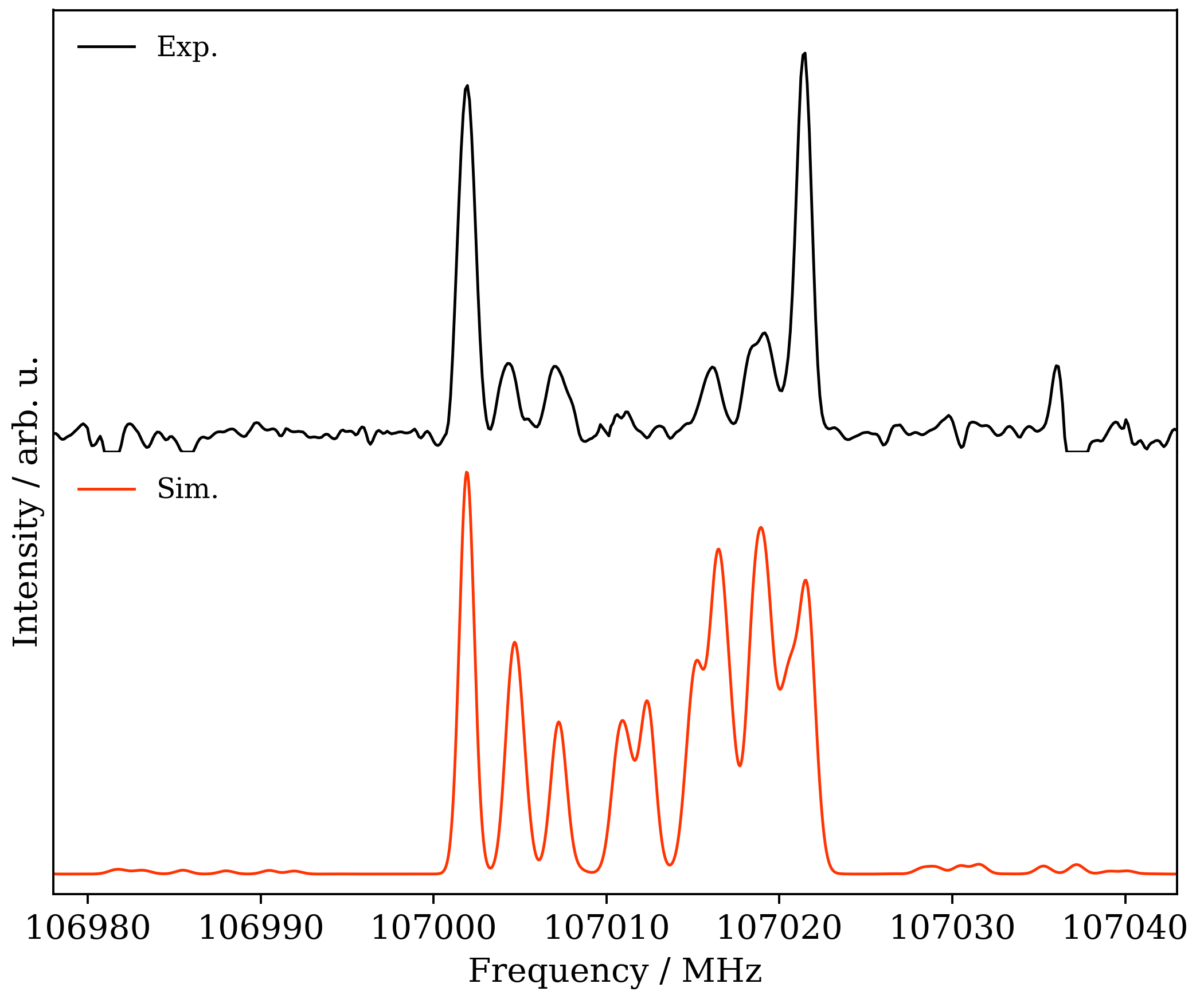}
    \caption{Experimental CP spectrum in the 107 GHz region where the $N'_{K'_a,K'_c}-N''_{K''_a,K''_c}=5_{0,5}-4_{0,4}$ $a$-type transition of \ch{H2NCO} lies, and comparison with a simulation obtained with PGOPHER using the final set of spectroscopic parameters (see Table~\ref{tab:param}), a 300\,K temperature, and a Gaussian profile with a full-width-at-half-maximum of 1 MHz. }
    \label{fig:hfs107}
\end{figure}

Following identification of \ch{H2NCO} on these spectra, experimental conditions were optimised to maximise the signal of this radical. A pulse length of 0.25\,\textmu s was found to maximise the signal of the observed $a$-type transitions, as expected from the relatively large projected dipole moment of 3.7\,Debye. Long acquisitions were subsequently undertaken, ultimately yielding a 2.9 million averaged FIDs spectrum. The procedure enabling to retrieve a radical-only spectrum was applied to that spectrum, the resulting spectrum is displayed in Fig.~\ref{fig:CPini} of the main text and a zoom onto the $N'_{K'_a,K'_c}-N''_{K''_a,K''_c}=5_{0,5}-4_{0,4}$ is shown on Fig.~\ref{fig:hfs107}.

\paragraph*{Other species synthesised}
On the final CP spectrum, many close-shell molecules synthesised in the experiment are visible. We observe notably strong lines of HCN (including transitions in $v_2=1$ and $v_3=1$, and of the H$^{13}$CN and HC$^{15}$N species), HNC, HNCO (including strong vibrational satellites and both $^{13}$C and $^{15}$N isotopologues), as well as lines of fluorinated species HFCO, \ch{FC(O)NH2}, \ch{F2CO}, and FCN (in $v=0$, $v_2=1$).
Transitions of \ch{FC(O)NH2} are observed for the first time in the 75--110\,GHz spectral region, see Appendix \ref{app:fconh2} for additional information.

\FloatBarrier

\section{Hyperfine structure and fit detail }\label{app:details}

\begin{figure}[ht!]
    \centering
    \includegraphics[width=\columnwidth]{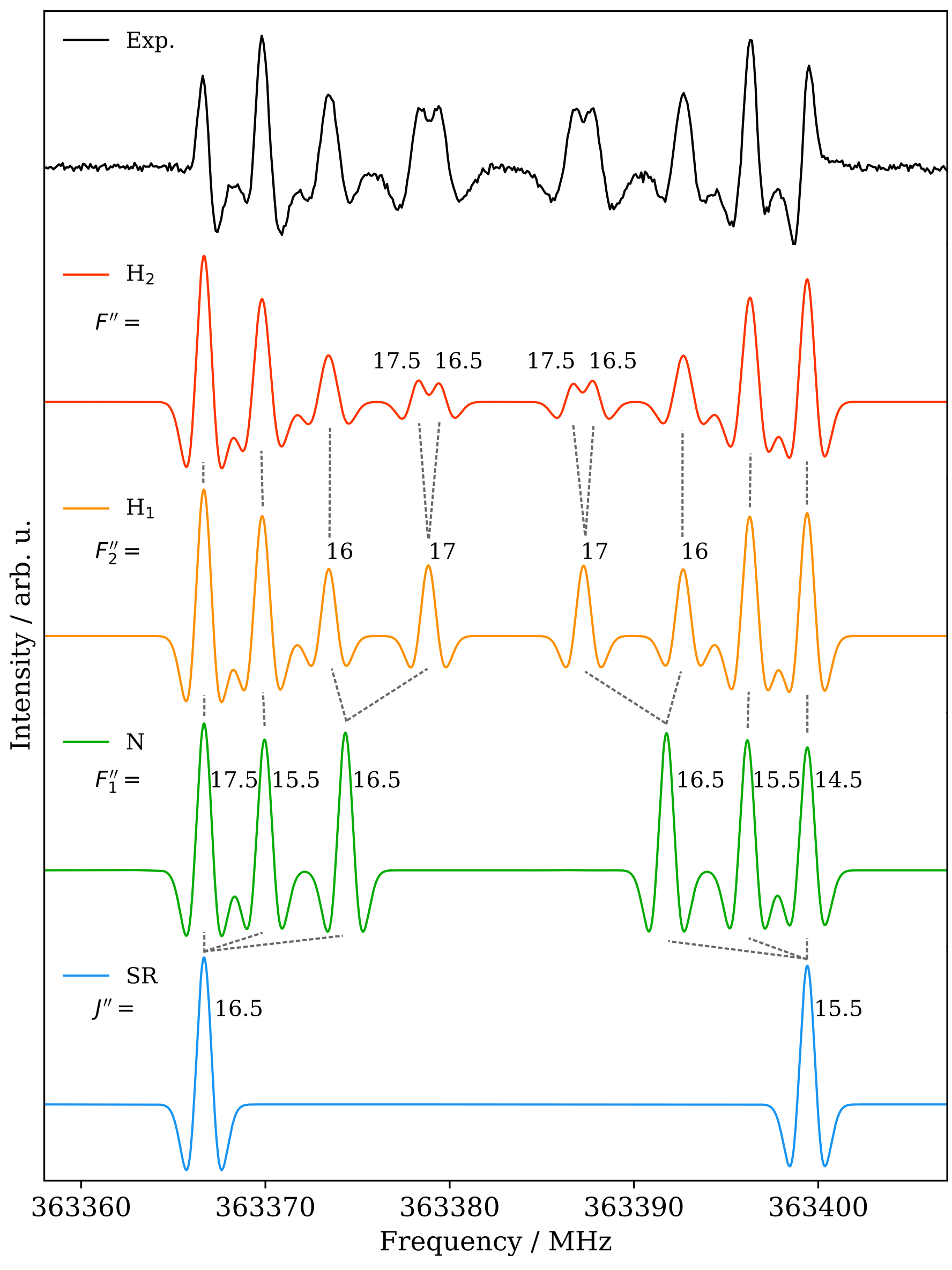}
    \caption{Evolution of the structure of the $N'_{K'_a,K'_c}-N''_{K''_a,K''_c}=17_{2,16}-16_{2,15}$ transition of \ch{H2NCO} upon consideration of the spin-rotation (SR) and hyperfine structures (N, H$_1$, and H$_2$), and comparison with the experimental spectrum. The quantum numbers of the different components are reported (unresolved values are omitted). The various simulations have been obtained the final spectroscopic constants (Table~\ref{tab:param}), a 300\,K temperature, and a Gaussian profile with a full-width-at-half-maximum of 1.3 MHz; the second derivative of the PGOPHER traces is plotted for visual comparison with the experimental spectrum. }
    \label{fig:hfs363}
\end{figure}

As mentioned in the main text, many transitions present a resolved hyperfine structure as illustrated in Fig.~\ref{fig:hfs363} where the experimental spectrum of the  $N'_{K'_a,K'_c}-N''_{K''_a,K''_c}=17_{2,16}-16_{2,15}$ transition is compared to simulation  of the fine structure ($J$ quantum number) and successive hyperfine structures ($F_1$, $F_2$, $F$ quantum numbers). On this transition, the central frequency components show a fully resolved hyperfine structure while the lowest and highest frequency components lie essentially at the frequency of nitrogen-hyperfine-split components. 
Special care was taken to include these components in the fit. 
To limit the number of lines in the fit, we have introduced two dummy vibrational states in the model (i.e., at the same energy as $v=0$), accounting for various cases of resolved hyperfine structure.
The fully resolved hyperfine structure corresponds to $v=0$ (red trace on Fig.~\ref{fig:hfs363}). For this state, all the parameters from Table~\ref{tab:param} are included.
Transitions displaying only N and H$_1$ hyperfine structure (orange trace on Fig.~\ref{fig:hfs363}) are coded as $v=1$; the hyperfine parameters pertaining to H$_2$ are omitted.
Finally, transitions only showing a N-resolved hyperfine structure (green trace on Fig.~\ref{fig:hfs363}) are coded as $v=2$; the hyperfine parameters pertaining to both H$_1$ and H$_2$ are omitted.
However, the hyperfine splitting is often asymmetrical with respect to the unsplit transition. A striking example is visible on Fig.~\ref{fig:hfs363} for the $F_1''=16.5 \rightarrow F_2''$ transitions but to some extent this also affects the $F_1''=15.5 \rightarrow F_2''$ transitions. Thus, particular care was taken in the choice of the hyperfine-case for the assignment. Concretely, an unresolved-case was only chosen if it did not induce a significant frequency shift for the transition.

Finally, Fig.~\ref{fig:hfs363} also illustrates the asymmetry in line profile observed for many transitions recorded by Zeeman-modulation spectroscopy, here the lowest and highest frequency components of the transition that exhibit a shape closer to that of a second derivative. This effect has been investigated in greater detail in Appendix~\ref{app:profile}.

\FloatBarrier
\section{Line profile in Zeeman modulation }\label{app:profile}

\begin{figure}[ht!]
    \centering
    \includegraphics[width=\linewidth]{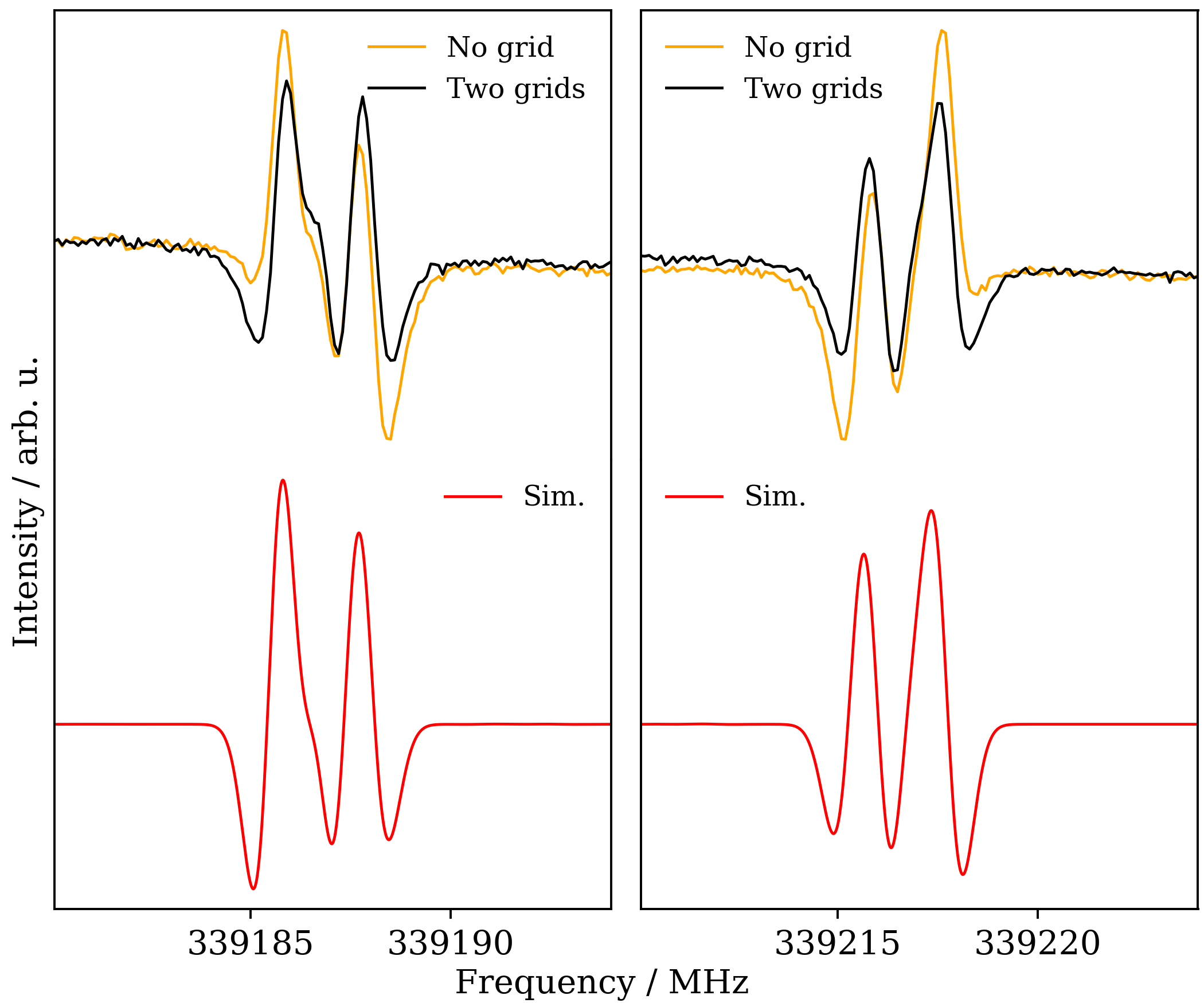}
    \caption{Experimental spectra of the $N'_{K'_a,K'_c}-N''_{K''_a,K''_c}=16_{0,16}-15_{0,15}$ transition of \ch{H2NCO} recorded without (in orange) and with (in black) polarisation grids, and comparison with a 300\,K simulation obtained with PGOPHER using the final set of spectroscopic parameters (see Table~\ref{tab:param}) and a Gaussian profile with a full-width-at-half-maximum of 1\,MHz. }
    \label{fig:grid}
\end{figure}

In the course of our measurements by Zeeman-modulation spectroscopy, we have noticed that some transitions exhibit an asymmetric line profile (e.g., see Figs.~\ref{fig:broad385} and \ref{fig:hfs363}). 
Some of these asymmetry may be explained by the effect of Faraday rotation of the polarisation around frequencies of radical transitions. We were able to partially correct this effect by adding polarisation grids in front of the source and detector, as illustrated in Fig.~\ref{fig:grid}.

\begin{figure}[ht!]
    \centering
    \includegraphics[width=\linewidth]{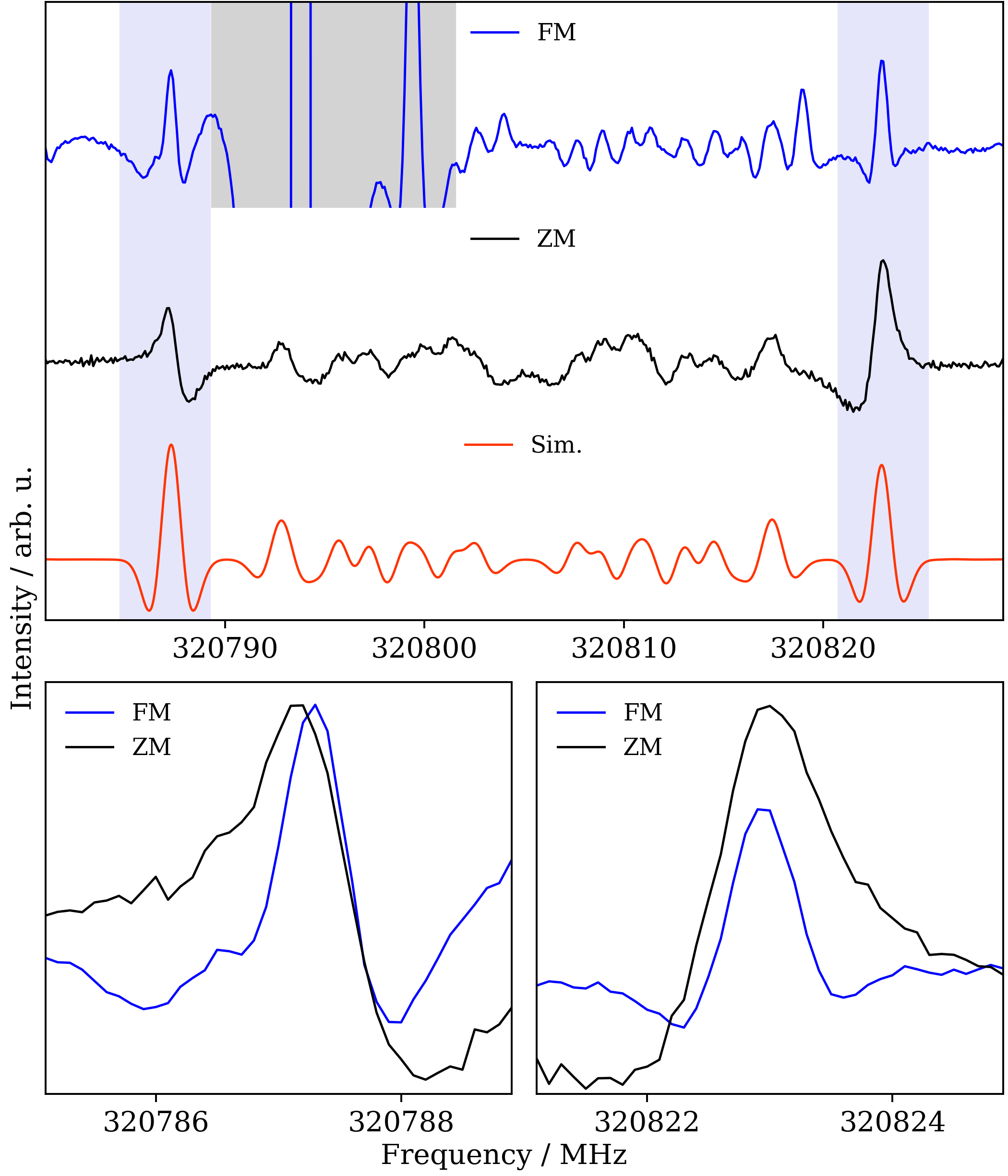}
    \caption{Experimental spectra recorded around the $N'_{K'_a,K'_c}-N''_{K''_a,K''_c}=15_{2,14}-14_{2,13}$ $a$-type transition of \ch{H2NCO} using source-frequency-modulation spectroscopy (FM) and Zeeman-modulation spectroscopy (ZM), and comparison with a 300\,K simulation obtained with PGOPHER using the final set of spectroscopic parameters (see Table~\ref{tab:param}) and a Gaussian profile with a full-width-at-half-maximum of 1.5\,MHz. The bottom panels show an overlap of the experimental spectra in the region highlighted in light blue on the top panel. The gray area on the upmost panel indicates a region where strong lines of the precursor prevent from observation of the radical product using the source-frequency-modulation experiment.}
    \label{fig:hfs320}
\end{figure}

For other transitions, the asymmetry of the transitions remained visible despite the use of the polarisation grids. 
This is striking for $K_a=2$ transitions (the lowest and highest frequency components are systematically affected). 
To better understand this effect, we have recorded the $N'_{K'_a,K'_c}-N''_{K''_a,K''_c}=15_{2,14}-14_{2,13}$ $a$-type transition using conventional source-frequency-modulation spectroscopy. The spectrum was acquired using a 48.157\,kHz source frequency modulation (FM), a 360\,kHz FM deviation, 100\,kHz frequency steps, and an acquisition time of 150\,ms per frequency point. The resulting spectrum is plotted on Fig.~\ref{fig:hfs320} where it is compared with the spectrum recorded using Zeeman modulation. A portion of the spectrum on the FM trace is not exploitable because of saturation of the precursor transitions, a common caveat when investigating discharge products using source-frequency-modulation spectroscopy. Other weaker transitions arising from close-shell species are also visible (for instance around 320819\,GHz). The acquisition using Zeeman-modulation strongly benefits from the open-shell species selectivity, as already mentioned in \citet{chahbazian2024:ch2cho}, although at the expense of the resolution since it yields broader features (the  $K_a=2$ transitions are particularly affected). 
By looking more closely at the two asymmetric components of the transition shown on Fig.~\ref{fig:hfs320} (lower panels), we observe that the maximum of the positive component of the line does not correspond to the centre frequency of the transition. Special care was thus taken to retrieve experimental frequencies for these components, and a large frequency uncertainty (typically 300\,kHz) was used in the fit.

Presently, we do not fully understand the line profile observed in Zeeman-modulation, nor the asymmetry of some of the components. It is interesting to note that the transitions particularly affected by the asymmetry already present some asymmetry in source-frequency-modulation spectroscopy, as visible in the lower panels of Fig.~\ref{fig:hfs320}, an effect that cannot be explained by the presence of unresolved hyperfine components. It is possible that these transitions are particularly affected by an external magnetic field (the Earth one in the case of the source-frequency-modulation spectrum). This would also be consistent with the fact that these transitions are overall stronger than other transitions on the Zeeman-modulation spectrum (as visible in Fig~\ref{fig:broad385}).

\FloatBarrier
\section{Vibrational satellites }\label{app:vibsat}

\begin{figure}[ht!]
    \centering
    \includegraphics[width=\linewidth]{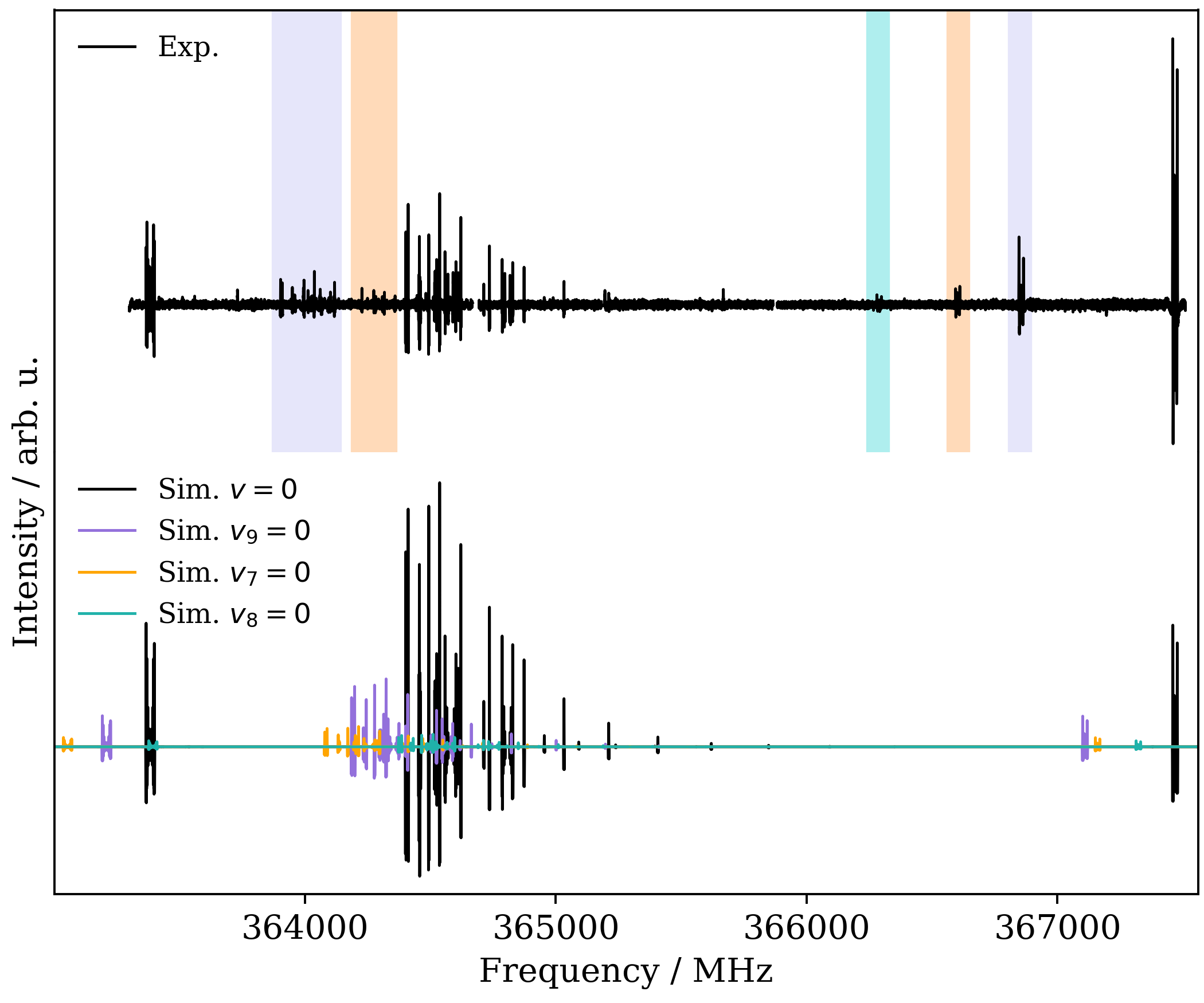}
    \caption{Experimental spectrum recorded around the $N'-N''=17-16$ $a$-type transition of \ch{H2NCO} (using Zeeman-modulation spectroscopy) (upper trace), and comparison with PGOPHER simulations for $v=0$ and the three lowest vibrationally excited states at 300\,K. The simulation in $v=0$ has been obtained using the final set of spectroscopic parameters (see Table~\ref{tab:param}). Those in the excited states have been obtained using the harmonic vibrational energies, the predicted rotational constants in these states (calculated using the $\alpha$ values in Table~\ref{tab:alphas}), and the other spectroscopic parameters from $v=0$. For all simulations, a Gaussian profile with a full-width-at-half-maximum of 1.3 MHz has been used. Shaded areas in the upper trace highlight regions where the observed transitions likely arise from vibrational satellites. }
    \label{fig:vibsat}
\end{figure}

The fundamental modes of \ch{H2NCO} calculated in this study are reported in Table~\ref{tab:vib}. At room temperature, the three lowest energy vibrational states $v_9=1$ ($\sim 282$\,cm$^{-1}$), $v_7=1$ ($\sim 540$\,cm$^{-1}$), and $v_8=1$ ($\sim 631$\,cm$^{-1}$) should possess significant population: \rev{13 to 26}\,\% (based on the \rev{anharmonic and harmonic} calculated energies), $\sim 7$\,\%, and $\sim 5$\,\%  of the ground state population, respectively. Indeed, on broad surveys of the experimental spectrum recorded using the Zeeman-modulation spectrometer (in particular in the 300--400\,GHz range were we have a very good signal-to-noise ratio), we observe several clusters of similar shape of those of \ch{H2NCO} in $v=0$, weaker and at lower frequencies. This is exemplified on Fig.~\ref{fig:vibsat} where the most prominent clusters have been highlighted using three shading colours. The lines in the light purple regions exhibit an intensity of about 25\,\% that of \ch{H2NCO}; they thus likely correspond to pure rotation in $v_9=1$. The clusters in the orange regions are about 10 times weaker than \ch{H2NCO} an thus may arise from transitions in $v_7=1$. Finally, a weak cluster is visible in the cyan region, with transitions intensities about 97\,\% weaker than \ch{H2NCO}; they may correspond to either $v_8=1$ or $v_9=2$.

Unambiguous identification of vibrational satellites requires spectroscopic assignments in these excited vibrational states. Simulations of the pure rotational spectrum of \ch{H2NCO} in $v_9=1$, $v_7=1$, and $v_8=1$ are presented in the lower panel of Fig.~\ref{fig:vibsat}, alongside the final simulation in $v=0$. The simulations have been obtained using the predicted rotational constants in the excited states (using the $\alpha$ values from our calculations, Table~\ref{tab:alphas}); all the other parameters being kept fixed to those in $v=0$. The harmonic energies of the excited states have been used since it was found to better reproduce the relative intensity of the transitions we believe are from $v_9=1$.
The most convincing identification is perhaps $v_9=1$, with a prediction that appears only slightly shifted from the clusters observed in the experimental spectrum. Individual transition assignments, however, have proven challenging based on this portion of the spectrum only. Indeed, we have not performed systematic surveys over the entire spectral region, and even for the $N'-N''=17-16$ transition shown in the figure, the experimental measurements lack coverage in the regions where the $K_a=0, 1$  and the lowest frequency $K_a=2$ transitions lie.
Dedicated measurements should be undertaken to fully unveil the spectrum of these vibrational satellites.

As a side note, Fig.~\ref{fig:vibsat} also illustrates the somewhat anomalous intensities of the $K''_a=2$ $a$-type clusters on the experimental spectrum. These clusters indeed appear much stronger than what expected from the simulation, an effect that is visible over the entire submillimetre spectral region probed in this study. It is worth noting that these $K''_a=2$ clusters are also those that exhibit the most asymmetric line profile. It seems that they are the most sensitive to the modulation of the magnetic field (as mentioned already in section \ref{app:profile}).

\FloatBarrier

\section{Partition function }\label{app:partFunc}

The \rev{rotational} partition function has been calculated at various temperatures using \texttt{SPCAT} for the fully resolved hyperfine structure (since components are observed over the entire submillimetre-wave domain). \rev{Special care has been taken to ensure convergence of the partition function (which required calculation up to $J=130$).}
\rev{Because the molecule possesses low-lying vibrational modes that are significantly populated at 300\,K as seen in Appendix~\ref{app:vibsat}, the vibration partition function has been calculated as well using the formula \citep{gordyMicrowaveMolecularSpectra1984}:
\begin{equation}
    Q_\mathrm{vib} = \Pi_{i=1}^{N_\mathrm{vib}} \frac{1}{1-e^{-E_i(\mathrm{vib})/kT}}
\end{equation}
where $N_\mathrm{vib}$ are the number of vibrational modes considered in the product, $E_i(\mathrm{vib})$ is the vibrational energy of the $i$th vibrational mode (in cm$^{-1}$, see Table \ref{tab:vib}), $k$ is the Boltzmann constant (expressed in cm$^{-1}$/K for the sake of homogeneity), and $T$ the temperature in Kelvin. Based on the results of the relative intensities of the $v_9=1$ vibrational satellites of \ch{H2NCO} observed on the experimental spectrum, that seem in agreement with the expected population assuming the harmonic energy if this mode, we decided to compute the harmonic value of the vibrational partition function ($Q_\mathrm{vib}=Q_\mathrm{vib}^\mathrm{harm}$).
For the calculation, we only considered the vibrational modes that holds significant population at room temperature, hence those with energies lower than 1000\,cm$^{-1}$. Namely, these consist in $v_9=1$, $v_9=2$, $v_9=3$, $v_7=1$, $v_8=1$, $(v_9,v_7)=(1,1)$, and $(v_9,v_8)=(1,1)$.
To total partition function is then obtained using the formula:
\begin{equation}
    Q_\mathrm{tot} = Q_\mathrm{rot} Q_\mathrm{vib}
\end{equation}
The different values are reported in Table~\ref{tab:partFunc}. 
}

\rev{The catalogue intensities values reported in the electronic supporting information have been calculated using $Q_\mathrm{tot}(300\,K)$. Intensities at lower temperatures can be recalculated using the corresponding $Q_\mathrm{tot}(T)$ values.} 

\begin{table}[ht!]
    \centering
    \caption{\rev{Rotational, vibrational, and total partition functions} for \ch{H2NCO} (fully resolved hyperfine structure) at various temperatures.}
    \begin{tabular}{rrrr}
    \toprule
\multicolumn{1}{c}{$T$ /\,K} & \multicolumn{1}{c}{\rev{$Q_\mathrm{rot}$}} & \multicolumn{1}{c}{\rev{$Q_\mathrm{vib}$}} & \multicolumn{1}{c}{\rev{$Q_\mathrm{tot}$}} \\ \midrule
    300.000 &  175482.9765 & 1.7259 & 302864.9020 \\
    225.000 &  113906.6451 & 1.3102 & 149235.8570 \\
    150.000 &   61973.9769 & 1.0860 & 67304.9490  \\
     75.000 &   21916.3156 & 1.0045 & 22016.0110  \\
     37.500 &    7761.8791 & 1.0000 & 7762.0343   \\
     18.750 &    2755.3989 & 1.0000 & 2755.3989   \\
      9.375 &     982.4627 & 1.0000 & 982.4627    \\
      \bottomrule
    \end{tabular}
    \label{tab:partFunc}
\end{table}

\FloatBarrier
\section{\ch{FC(O)NH2}} \label{app:fconh2}

\begin{figure}[ht!]
    \centering
    \includegraphics[width=\linewidth]{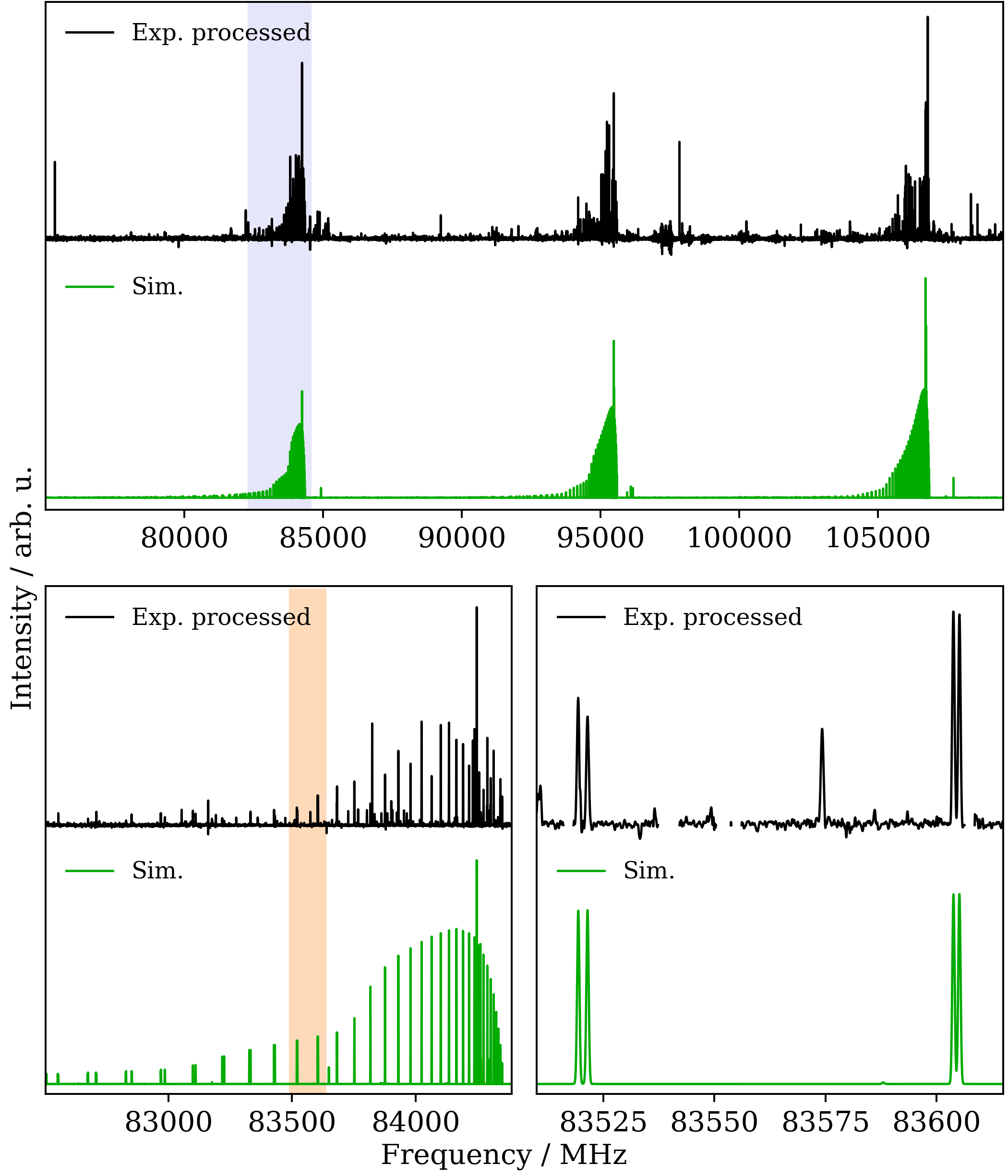}
    \caption{CP experimental spectrum after processing allowing to remove contributions from known close-shell species (upper trace), and comparison with a PGOPHER simulation for \ch{FC(O)NH2} in $v=0$. The simulation has been obtained using the final set of spectroscopic parameters (see Table~\ref{tab:paramFCONH2}), a temperature of 300\,K, and a Gaussian profile with a full-width-at-half-maximum of 0.6 MHz. The shaded area in light purple in the upper panel is zoomed-in the bottom left panel; that in orange is zoomed-in in the lower right panel. Discontinuities in the processed experimental spectrum correspond to region where lines of other molecules were lying.}
    \label{fig:fconh2}
\end{figure}

\begin{table}
    \centering \scriptsize
    \caption{Spectroscopic parameters of \ch{FC(O)NH2} in the $S$-reduction and relevant fit information. 1$\sigma$ error are reported within parentheses in the unit of the last digit; they have been formatted using the PIFORM software developed by \citet{kisiel2001:Assignment}.}
    \label{tab:paramFCONH2}\footnotesize
    \begin{tabular}{ll d{5.2}d{5.8}}
\toprule
Parameter         & units & \multicolumn{1}{c}{Lit.$^a$} & \multicolumn{1}{c}{Exp. $v=0$}  \\
\midrule
$A$                                 & /MHz &   11304.39  &   11304.4357(40)  \\
$B$                                 & /MHz &   11176.14  &   11176.1677(37)  \\
$C$                                 & /MHz &   5616.06   &    5616.1001(32)  \\
$D_J$                               & /kHz &             &       5.078(29)   \\
$D_{JK}$                            & /kHz &             &       3.571(35)   \\
$D_K$                               & /kHz &             &       2.576(36)   \\
$d_1$                               & /kHz &             &      -3.1269(30)  \\
$d_2$                               & /kHz &             &      -0.9523(19)  \\
$H_{JK}$                            & /Hz  &             &       0.312(14)   \\
$H_{KJ}$                            & /Hz  &             &      -0.428(26)   \\
$H_{K}$                             & /Hz  &             &       0.185(20)   \\
$h_2$                               & /Hz  &             &       0.01144(58) \\
\\              
$3/2\chi_{aa}($N$)$                 & /MHz &             &      3.63(14)     \\
$1/4(\chi_{bb}-\chi_{cc})($N$)$     & /MHz &             &      1.484(34)    \\
\midrule

\multicolumn{2}{l}{$\# / n\,^{b}$}
& \multicolumn{1}{c}{$96 / 46$}
& \multicolumn{1}{c}{$1181 + 96 / 215 + 46$}
\\
\multicolumn{2}{l}{$J'_\mathrm{max},\,K'_{a\,\mathrm{max}}$}
& \multicolumn{1}{c}{$7,\,6$} 
& \multicolumn{1}{c}{$50,\, 42$} 
\\
\multicolumn{1}{l}{RMS $\,^c$} & /MHz
& \multicolumn{1}{c}{}
& \multicolumn{1}{c}{0.102} 
\\
\multicolumn{2}{l}{$\sigma\,^{d}$}
& \multicolumn{1}{c}{}
& \multicolumn{1}{c}{1.02} 
\\
\bottomrule     
    \end{tabular}
    
\smallskip

\begin{minipage}{0.9\columnwidth}  \footnotesize
    $^a$ \cite{rigden1966:Microwave}\\    
    $^b$ Number of lines ($\#$) and number of different frequencies ($n$) \\
    $^c$ Root mean square value\\
    $^d$ Weighted standard deviation
\end{minipage}
\end{table}

Upon assignment of transitions from known molecules on the CP-FTMMW spectrum, we have observed three clusters of transitions, lying around 84, 95, and 106\,GHz, as shown in Fig.~\ref{fig:fconh2}. These lines were rapidly assigned to \ch{FC(O)NH2} for which transitions in $v=0$ have previously been measured in the 16--40\,GHz range by \citet{rigden1966:Microwave} using Stark spectroscopy  ($J' \leq 4\,, K_a' \leq 6$, 46 different frequencies). A fit of the literature transitions with a Watson-$S$ Hamiltonian using the Pickett's SPFIT software has enabled spectral predictions in the W-band range and assignments of the observed transitions of that molecules.  We assigned 215 lines (both $a$- and $b$-type) corresponding to transitions involving $J' \leq 50$ and $K_a' \leq 42$. Many of the transitions exhibit a partially resolved nitrogen hyperfine structure. Because of unresolved asymmetric splitting for most of the transitions, the total number of lines in the fit is much larger (1277 total). Transitions from \citet{rigden1966:Microwave}  and the CP-FTMMW spectrum were assigned a 100--200\,kHz and 50--150\,kHz frequency uncertainty, respectively.
The resulting spectroscopic parameters are reported in Table~\ref{tab:paramFCONH2} and a simulation using these constants is shown in Fig.~\ref{fig:fconh2}. The corresponding fit files are available in the electronic supplementary material.

\FloatBarrier
\section{\rev{\ch{HNCHO}}} \label{app:hncho}

\begin{figure}
    \centering
    \includegraphics[width=1.0\columnwidth]{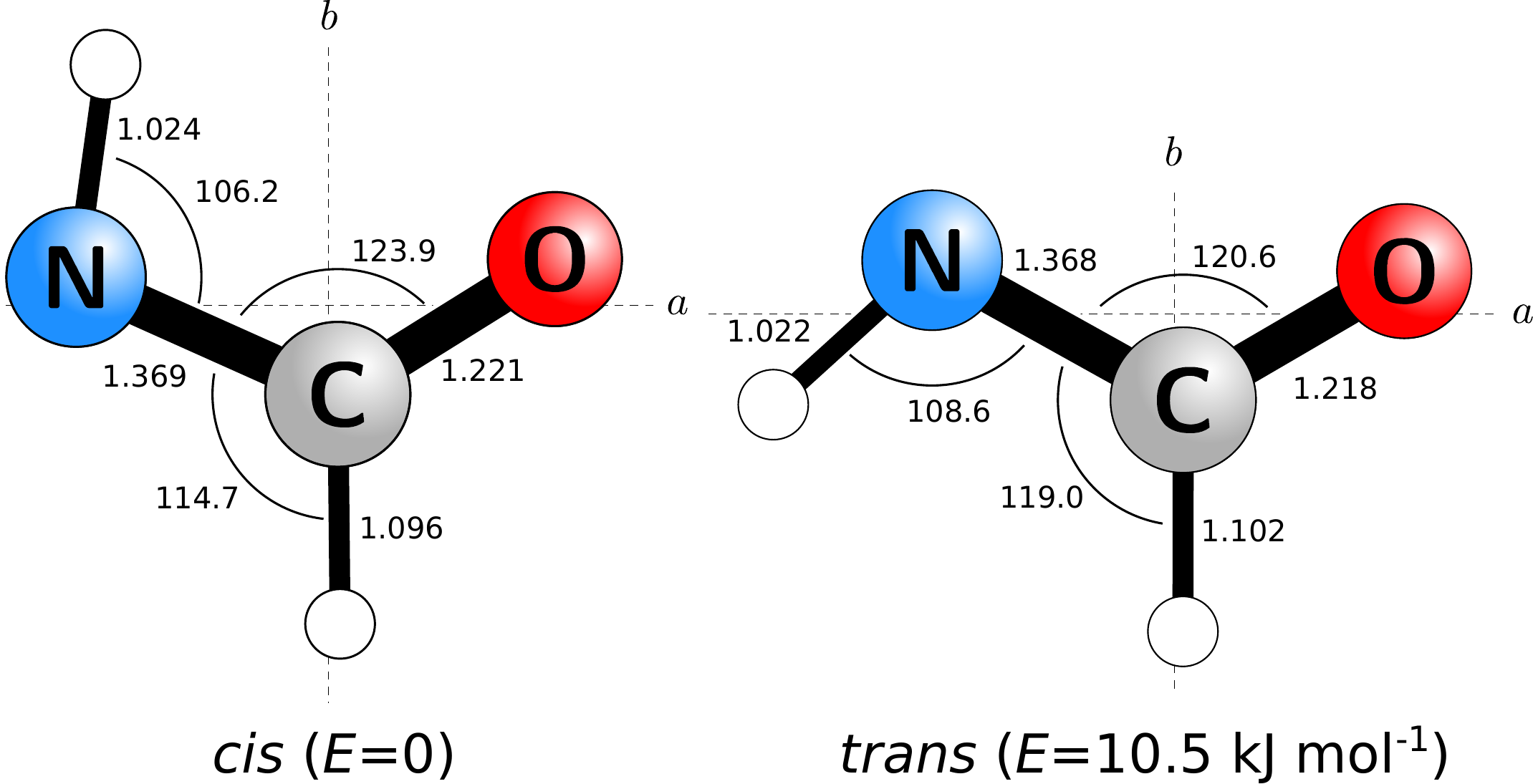}
    \caption{Equilibrium structure of \textit{cis} and \textit{trans} \ch{HNCHO} calculated at the $\omega$B97X-D/cc-pVQZ level of theory. Geometrical parameters (bond lengths in angstroms and angles in degrees) and principal axes of inertia are reported. The figure was generated using the PMIFST software from the PROSPE collection of programs developed by \citet{kisiel2001:Assignment}. }
    \label{fig:structisom}
\end{figure}

HNCHO was investigated by quantum-chemical calculations in a similar fashion as its \ch{H2NCO} isomer at the $\omega$B97X-D/cc-pVQZ level of theory. The radical possesses two planar conformations, both belonging to the $C_s$ point group of symmetry; namely \textit{cis} (H--N--C--O dihedral angle of 0 \textdegree) and \textit{trans} (H--N--C--O dihedral angle of 180 \textdegree), the latter lying 10.5 kJ/mol higher than the former. Both conformer are displayed in Fig. \ref{fig:structisom} along with relevant structural information. The permanent dipole moment projections take values of $\mu_a =0.7$ Debye and $\mu_b =0.2$ Debye for the \textit{cis} conformer and $\mu_a =2.0$ Debye and $\mu_b =2.7$ Debye for the \textit{trans} one. The predicted spectroscopic constants for these two conformers are reported in Table \ref{tab:paramisom}.

\begin{table}
    \centering \scriptsize
    \caption{Calculated spectroscopic parameters (in MHz) of \ch{HNCHO} \textit{cis} and \textit{trans} conformers in the $S$-reduction.}
    \label{tab:paramisom}\footnotesize
    \begin{tabular}{l d{5.6}d{5.6}}
\toprule
Parameter          & \multicolumn{1}{c}{\textit{cis}} & \multicolumn{1}{c}{\textit{trans}}  \\
\midrule
$A$                                 &   72610.71      &   81883.12      \\
$B$                                 &   12407.62      &   12028.24      \\
$C$                                 &   10596.84      &   10487.66      \\
$D_N$                               &   0.01003       &   0.007903      \\
$D_{NK}$                            &   0.02089       &   0.09529       \\
$D_K$                               &   1.256        &   1.962         \\
$d_1$                               &   -0.002003     &   -0.001256     \\
$d_2$                               &   -0.000264     &   -0.000230     \\
\\                                                          
$\epsilon_{aa}$                     &   -2388.        &     -2170.       \\
$\epsilon_{bb}$                     &   -109.4        &     -100.1       \\
$\epsilon_{cc}$                     &   3.848         &     3.910        \\
   \\                                                       
$a_F($N$)$                          &  14.23          &     13.21        \\
$3/2T_{aa}($N$)$                    &  -52.84         &     -51.24       \\
$1/4(T_{bb}-T_{cc})($N$)$           &  -26.17         &     -26.12       \\
$T_{ab}($N$)$                       &  -0.3635        &     0.1979       \\
$3/2\chi_{aa}($N$)$                 &  -6.366         &     0.4001       \\
$1/4(\chi_{bb}-\chi_{cc})($N$)$     &  -1.523         &     -2.569       \\
$\chi_{ab}($N$)$                    &  -2.542         &     0.7746       \\
\\                                                          
$a_F($H$_1)$                        &   4.403         &     5.531       \\
$3/2T_{aa}($H$_1)$                  &   7.770         &     7.732       \\
$1/4(T_{bb}-T_{cc})($H$_1)$         &   2.032         &     2.017       \\	
$T_{ab}($H$_1)$                     &   -5.602        &     -4.618      \\
\\                                                                    
$a_F($H$_2)$                        &   -50.29        &   -51.25         \\
$3/2T_{aa}($H$_2)$                  &   -60.65        &   15.19          \\
$1/4(T_{bb}-T_{cc})($H$_2)$         &   13.90         &   0.2089         \\	
$T_{ab}($H$_2)$                     &   9.995         &   47.63          \\

\bottomrule     
    \end{tabular}
    
\smallskip

\begin{minipage}{1\columnwidth}  \footnotesize
    $^a$ The calculated $A, B, C$ rotational constants have been Bayesian corrected by a 0.9866 factor \citep[see][]{lee2020:Bayesian}, all the other calculated parameters are equilibrium values.   
\end{minipage}
\end{table}

\end{appendix}
\end{document}